\documentclass[]{aa}  

\usepackage{graphicx}
\usepackage{txfonts}
\usepackage{natbib}

\usepackage{color}
\usepackage{textcomp,gensymb}
\usepackage{booktabs}
\usepackage{amssymb}
\usepackage{amsfonts}
\usepackage{amsmath}
\usepackage{mathrsfs}
\usepackage{wasysym}

\newcommand{\dd}{\mathrm{d}}
\renewcommand{\d}{\mathrm{d}}
\newcommand{\Pp}{\mathscr{P}}

\newcommand{\eps}{\epsilon}
\newcommand{\di}{\displaystyle}

\newcommand{\corr}[1]{{#1}}
\newcommand{\corrtwo}[1]{{#1}}

\begin{document}

   \title{A stratified jet model for AGN emission in the two-flow paradigm.}

  \author{T. Vuillaume
          \inst{1}
          \and
          G. Henri \inst{2}
          \and
          P-O. Petrucci \inst{2}
          }

   \institute{\inst{1}Univ. Grenoble Alpes, Univ. Savoie Mont Blanc, CNRS, LAPP, 74000 Annecy, France\\
   \inst{2}Univ. Grenoble Alpes, CNRS, IPAG, 38000 Grenoble, France \\
              \email{thomas.vuillaume@lapp.in2p3.fr}
             }

   \date{}

\abstract{\corr{High-energy emission of extragalactic objects is known to take place in relativistic jets, but the nature, the location, and the emission processes of the emitting particles are still unknown. One of the models proposed to explain the formation of relativistic ejections and their associated non-thermal emission is the two-flow model, where the jets are supposed to be composed of two different flows, a mildly relativistic baryonic jet surrounding a fast, relativistically moving electron positron plasma. Here we present the simulation of the emission of such a structure taking into account the main sources of photons that are present in active galactic nuclei (AGNs). }}{\corr{We try to reproduce the broadband spectra of radio-loud AGNs with a detailed model of emission taking into account synchrotron and inverse-Compton emission by a relativistically moving beam of electron positron, heated by a surrounding turbulent baryonic jet.}}{\corr{We compute the density and energy distribution of a relativistic pair plasma all along a jet, taking into account the synchrotron and inverse-Compton process on the various photon sources present in the core of the AGN, as well as the pair creation and annihilation processes. We use semi-analytical approximations to quickly compute the inverse-Compton process on a thermal photon distribution with any anisotropic angular distribution. The anisotropy of the photon field is also responsible for the bulk acceleration of the pair plasma through the "Compton rocket" effect, thus imposing the plasma velocity along the jet. As an example, the simulated emerging spectrum is compared to the broadband emission of 3C273.}}{\corr{In the case of 3C273, we obtain an excellent fit of the average broadband energy distribution by assuming physical parameters compatible with known estimates. The asymptotic bulk Lorentz factor is lower than what is observed by superluminal motion, but the discrepancy could be solved by assuming different acceleration profiles along the jet.}}{} 

   \keywords{Galaxies: jets --
                Galaxies: active --
                Radiation mechanisms: non-thermal -- gamma rays: theory -- quasars: individual: 3C273
               }

   \maketitle
%
%________________________________________________________________

\section{Introduction}
\label{intro}

Radio-loud active galactic nuclei (AGNs)  are known to be powerful emitters of non-thermal radiation. In the simplest leptonic models, this emission is most commonly attributed to the presence of highly relativistic leptons accelerated in a relativistically moving magnetized jet, emitting synchrotron radiation and inverse-Compton photons on various sources of soft photons. 
However several features are still unclear, such as the composition of the jet (leptonic $\rm e^+ / e^-$ or baryonic $\rm p^+ / e^-$ ), the bulk acceleration mechanism, the heating mechanism of the relativistic non-thermal particles, \corr{and} the precise size and location of the emitting zones.  One-zone models are the most simple and widespread models for reproducing the AGN jet emission. They assume a spherical, homogeneous emission zone with a minimal number of free parameters: the radius of the zone, the magnetic field, the bulk Lorentz factor, and parameters describing the particle distribution.  They present the advantage of being simple and give a good first approximation of the physical conditions in jets. However, they encounter several limitations. Among others, they have difficulties in reproducing the low-energy (radio) part of the spectral energy distribution (SED; most probably emitted by the farthest part of the jet).  They assume the very stringent condition that the whole non-thermal emission must be produced in a single zone, which can be an issue for explaining the multi-wavelength variability of the sources.  Furthermore, they do not offer any clue on the jet formation mechanism and its parameters outside this zone. \corr{Also the strong Doppler boosting associated with highly relativistic jets is incompatible with the detection of high-energy emission from unbeamed radio galaxies seen at large angles, since their emission should be strongly attenuated by a Doppler factor smaller than one.} \\ 

Facing these weaknesses, models \corr{considering more complex structures} have been proposed involving stratified  inhomogeneous jets. \corr{For instance, the b}lob-in-jet models \citep{Katarzyski:2001iga,Hervet:2015wo} propose a structure where blobs move at high relativistic speed in the jet. Those blobs can be responsible for some of the emission (especially at high energies) whereas the ambient jet can explain the rest of the spectrum, for example in the radio band. Spine/sheath models, implying the existence of two flows at different velocities,  can provide a picture in agreement with both theoretical and observational constraints. \cite{Ghisellini:2005bc} developed a model based on the same idea and showed that this kind of structure could help reduce the necessity for very high bulk Lorentz factor values as the two emitting zones interact radiatively, enhancing each other's emission.\\

But the original idea of a double jet structure stemmed from \cite{1989MNRAS.237..411S}, who coined the name "two-flow model" (more details in Sect. \ref{sec:twoflow}).
In their original paper, the authors proposed for the first time a double jet structure for AGN jets. \corr{In this model, an outer jet or collimated wind is ejected from an accretion disc, with a mildly relativistic velocity \corr{($\upsilon \sim0.5c$)}}. {In the empty funnel of this jet, a fast inner electron-positron beam is formed and moves at much higher relativistic speeds (bulk Lorentz factor $\Gamma_b \approx 10$). The pairs are supposed to be continuously heated by the surrounding baryonic jet through MHD turbulence.}\\

\corr{The model has several advantages compared to models assuming a single fluid. The first advantage is that the problem of the energy budget of relativistic jets is reduced, since only a minor component of leptons needs to be accelerated to high bulk Lorentz factors. The protons of the surrounding jets are not supposed to be highly relativistic}. The second is that this model can provide a simple way to solve the discrepancy between the required high Lorentz factors to produce the observed gamma-ray emission and the slower observed motion in jets at large scales (e.g. \citealt{2006ApJ...640..185H} and references therein). \corr{Furthermore, as the power is carried out mainly by the non-relativistic jet, it escapes the Compton drag issue. As explained below, the pair beam is only gradually accelerating thanks to the anisotropic inverse-Compton effect (or "Compton rocket" effect), and its velocity never exceeds the characteristic velocity above which the aberrated photon field starts to cause a drag (it actually remains at this characteristic velocity). Its density increases all along the jet due to the gamma-gamma pair production process, so the dominant emission region can be at large distances from the central black hole, avoiding the problem of gamma-gamma absorption by the accretion disk photon field. This model therefore offers a natural explanation of the main characteristics of the high-energy source deduced from observations.  } \\

Noticeably, the dynamical effects of radiation on relativistic particles (which are intrinsically strongly dissipative) are very difficult to incorporate both in analytical and in numerical (M)HD simulations. The picture we present here is thus markedly different from most models available in the literature, since the pair component dynamics is mainly governed (at least at distances relevant for high-energy emission) by these radiative effects. However we argue that these effects are unavoidable since the cooling time of relativistic leptons is indeed very short at these distances, and these effects must be taken into account in any physically relevant model implying a relativistic pair plasma, which is in turn likely to exist given the high density of gamma-ray photons observed from radio-loud gamma-ray-emitting AGNs.\\

The first numerical model of non-thermal emission based on these ideas was proposed by \cite{1995MNRAS.277..681M},\corr{who considered only the inverse-Compton process on accretion-disk photons.} Assuming  a power-law  particle distribution and a stratified jet, the authors could derive the inverse-Compton emission from a plasma of relativistic leptons illuminated by a standard accretion disc \corr{as well as the opacity to pair production and the pair production rate }. They showed that the spontaneous generation of a dense  $e^+ / e^-$  pair beam continuously heated by the baryonic jet was indeed possible and was able to reproduce \corr{the gamma-ray emission of EGRET blazars}.\\

Based on this work, \cite{1996A&AS..120C.563M}  further studied the possibility of accelerating the $e^+ / e^-$  pair beam in the framework of the two-flow through the Compton rocket mechanism introduced by \cite{1981ApJ...243L.147O}. In this mechanism, the motion of the relativistic pair plasma is entirely controlled by the local anisotropy of the soft photon field, which produces an anisotropic inverse-Compton emission transferring momentum to the plasma. The acceleration saturates when the photon field, aberrated by the relativistic motion, appears nearly isotropic (vanishing flux) in the comoving frame. 
\cite{1998MNRAS.300.1047R} continued this work on the acceleration via the Compton rocket effect. Considering a relativistic pair plasma following a power-law energy distribution coupled with the photon field from a standard accretion disc and extended sources (BLR or dusty torus); they computed the value of the terminal bulk Lorentz factor and showed that values up to \corr{$\Gamma_b = 20$} are achievable for extragalactic jets, the most probable values being of the order of 10  in good agreement with VLBI motions studies (e.g. \citealt{Lister:2009hn}).
Subsequent works studied the possibility of explaining the spectra by a pile-up (relativistic Maxwellian) distribution \citep{Sauge:2004ep}, which better reproduces the high-energy spectra of BL Lacs \corr{that peak in the TeV range. A quasi-Maxwellian distribution is close to a monoenergetic one, however the spatial convolution of such a distribution whose parameters vary along the jet can mimic a power-law over a limited range of frequencies.} These authors also proposed a time-dependent version of the model that was \corr{later shown to be } able to successfully reproduce the rapid flares observed in some objects, such as PKS 2155-304 \citep{Boutelier:2008bga}.\\

In a recent study, \cite{Vuillaume:2015jv} (hereafter Vu15)  further studied the acceleration through the Compton rocket effect. The complex photon field of an AGN was considered,  carefully taking into account the spatial distribution of extended sources (standard accretion disc, dusty torus, and broad line region). The evolution of the resulting bulk Lorentz factor along the jet is then computed self-consistently and it appears that due to the complexity of the surrounding photon field, it can display a relatively tangled relation with the distance to the base of the jet. Moreover, variations of the bulk Lorentz factor (and thus of the Doppler factor) can induce complex variability of the observed emission in space and in time.\\

The goal of this paper is to present the complete calculation of the pair-beam non-thermal spectra {\bf assuming} the above configuration, that is, 1) the pair plasma is assumed to be described by a pile-up distribution and is generated in situ by $\gamma-\gamma$ interactions, 2) its motion is controlled {at short and intermediate distances (up to $\sim 10^3 r_g$) by the anisotropic Compton rocket effect in the complex photon field generated by an accretion disk, a broad line region, and a dusty torus, and 3) particles emit non-thermal radiation through synchrotron, synchrotron self-Compton (SSC), and external Compton (EC) in the various photon fields. Part of the high-energy $\gamma-$ray photons can also be absorbed to produce new pairs. We assume that these new pairs are continuously accelerated along the jet. In the two-flow paradigm \citep{1989MNRAS.237..411S}, such acceleration is expected by the MHD turbulence generated by the surrounding baryonic flow. Subsequently, the density and internal energy of the pair plasma are computed self-consistently, and the emitted photon spectrum can be evaluated and compared to observations. \\

The overall layout is as follows. In Section \ref{sec:twoflow} we present the main theoretical interest of the two-flow paradigm. Then in Section \ref{sec:num_model} we detail our model and the numerical methods we use. In section \ref{sec:3c273} we apply this model to the \corr{bright} quasar 3C273 and show the model can reproduce its jet emission from radio to gamma-rays.

\section{The two-flow paradigm: the hypothesis and the reasoning behind it\label{sec:twoflow}}

The two-flow paradigm is based on an original idea from \citet{1989MNRAS.237..411S} (see Section \ref{intro}).The model has evolved since then but the core hypothesis remains the same: an AGN ``jet'' is actually made of two interacting flows.
The outer one is a MHD jet, or wind, fuelled by the accretion disc. It originates from and is self-collimated by the Blandford \& Payne (BP) process \citep{Blandford:1982vy} and is much like the jets found in other objects such as young stars or neutron stars. This baryonic jet is therefore mass loaded by the disc and mildly relativistic ($\beta \approx 0.5$). On the rotation axis, where the angular momentum tends to zero, no MHD force counteracts the gravity from the central object and it is expected that the density strongly decreases near the axis, thus leaving an empty funnel. A lighter jet made of relativistic leptons $\rm (e^-/e^+)$ can then be created there through $\gamma-\gamma$ interaction between the surrounding soft and high-energy photons. This leptonic plasma is accelerated through the Compton rocket effect (as explained below) and travels at highly relativistic speeds. It is assumed to be confined, collimated, and continuously reheated by the surrounding MHD jet.

\begin{figure}[ht]
\begin{center}
        \includegraphics[width=\hsize]{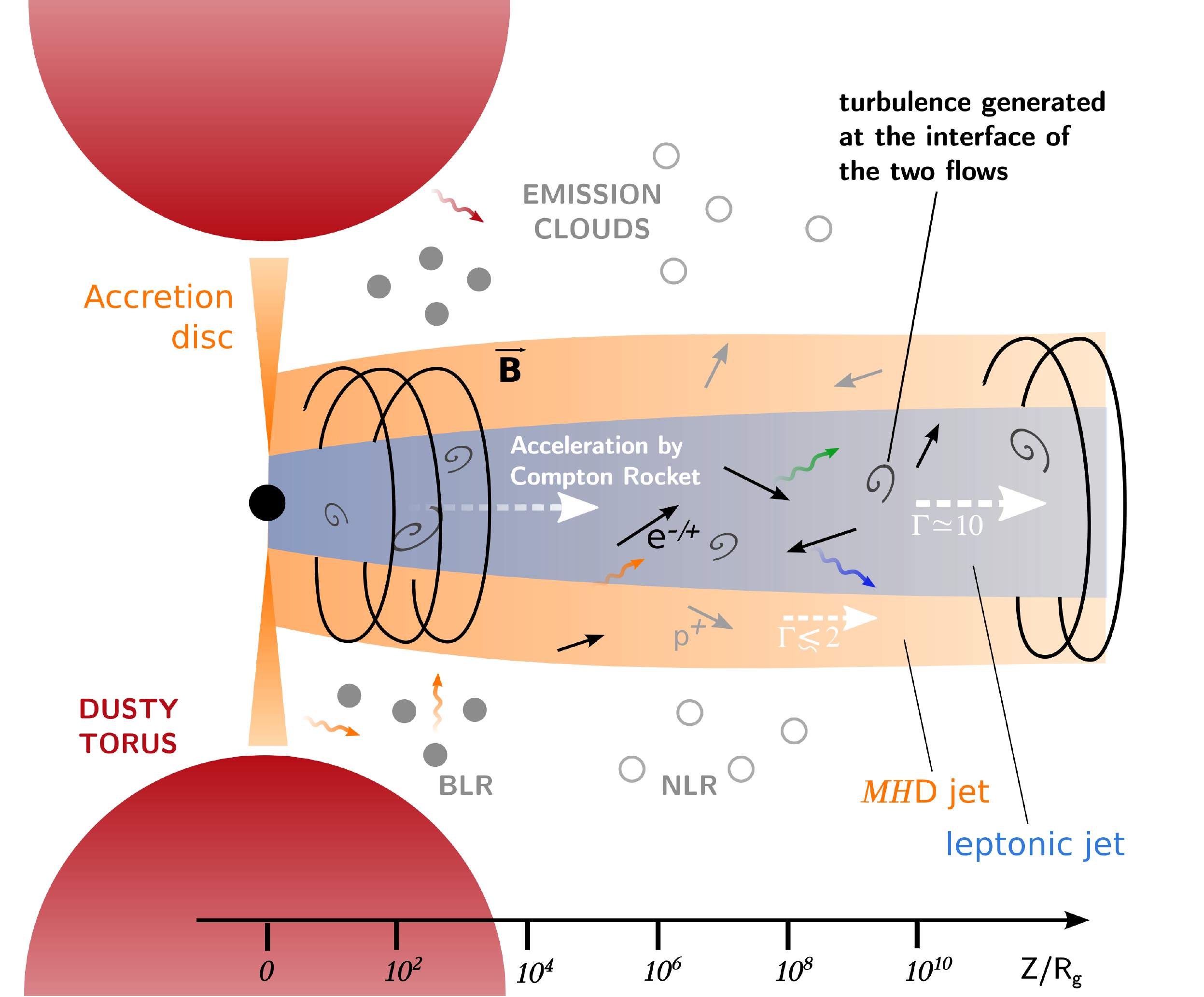}
\caption{Schematic view of the model developed in the two-flow paradigm.}
\label{fig:two-flow}
\end{center}
\end{figure}

\subsection{Interaction of highly relativistic flows}

The first benefit of the two-flow hypothesis is to alleviate the problem of the confinement of a highly relativistic flow. Self-confinement of a jet can take place due to the magnetic field exerting a magnetic pressure (from the Lorentz force) balancing internal pressure and centrifugal force. However, the self-confinement of a highly relativistic jet through this process is quite inefficient. This has been shown first in numerical simulations by \citealt{Bogovalov:2001jk} and \citealt{Bogovalov:2001ii} and then demonstrated based on theoretical arguments by \citealt{Pelletier:2004vj}.
%\footnote{the bottom line is that electric field increases in relativistic outflows and become too strong to allow self-collimation by the magnetic field}
Therefore the collimation of relativistic flows necessarily requires an external process like for example external pressure from the ambient (interstellar) medium.. In the two-flow paradigm, the outer self-confined massive and powerful MHD sheath confines the spine by ram pressure, providing an easy solution to the important problem of confinement of highly relativistic flows.
\newline

Moreover, the interface of the two flows can be subject to Kelvin-Helmholtz instabilities producing turbulence. This turbulence can then accelerate particles in the spine through second-order Fermi processes. In that picture, because the MHD sheath carries most of the power, it can be seen as an energy reservoir continuously energizing the particles through turbulence. 
This continuous source of energy gives rise to two very interesting phenomena.

The first one is the most important feature of the pair creation process. As new pairs are created through $\gamma-\gamma$ absorption, and immediately accelerated through turbulence to reach high energies, they can emit $\gamma$-rays that will create more pairs. The pair-creation process being very efficient and highly non-linear, a copious amount of new pairs can be created even from an initial, very low density, therefore constituting the spine. Moreover, above a certain threshold of energy, the process can runaway and give rise to episodes of rapid flares. This has been demonstrated by \cite{Renaud:1999vz}, \cite{Sauge:2004ep}, and \cite{Boutelier:2008bga}.\\

The second phenomenon is the possibility to accelerate the spine jet to relativistic motion through the anisotropic Compton rocket effect as discussed below.

\subsection{Bulk Lorentz factor of the spine\label{sec:jet_velocity}}
\corr{
The question of the actual speed, or bulk Lorentz factor $\Gamma_b$ of jets, as well as their acceleration, is central to the understanding of their physics. This is a long-standing and debated issue in the community \citep{2006ApJ...640..185H} with conflicts between theoretical arguments (\citealt{Aharonian:2007ep}, \citealt{Tavecchio:2010hc}, \citealt{2008MNRAS.384L..19B}) and observations \citep{Piner:2004cj, 2004ApJ...600..127G, Piner:2014io, Lister:2013gp}.
Some of the attempts to solve this issue come in the form of structured jets \citep{2000A&A...358..104C, Ghisellini:2005bc, 2003ApJ...594L..27G}.
}

In the two-flow paradigm, the question of the acceleration of the spine is solved by the Compton rocket effect as proposed by \cite{1981ApJ...243L.147O}. In this process, inverse-Compton scattering in a strongly anisotropic photon field induces a motion of the emitting plasma in the opposite direction. \cite{1981ApJ...243L.147O} showed already that a purely leptonic hot plasma must be dynamically driven by this process, up to relativistic speeds.
\cite{Phinney:1982wt} opposed the fact that this process is actually quite inefficient at accelerating a pair blob as the pairs cool down very quickly through the inverse-Compton scattering, therefore killing it before it can be effective.

However, in the two-flow paradigm, as the pairs are continuously re-accelerated by the turbulence all along the jet, this argument does not hold and the Compton rocket process becomes an efficient source of plasma thrust (see Section \ref{sec:gamma_b}). This effect is likely to be very efficient in type II AGNs (FSRQ and FR II galaxies) where an accretion disk with high luminosity is present. The ambient photon field in the first hundreds of $r_g$ will therefore be highly anisotropic and a pair plasma will be rapidly accelerated to a characteristic velocity for which the net aberrated flux, evaluated in the comoving frame, vanishes. As the cooling timescale for near-Eddington luminosities is much shorter than the other dynamical timescales, this effect will dominate over all other terms such as pressure gradients and magnetic effects. Hence the velocity of the plasma will be very close to this characteristic equilibrium velocity. If the photon field is mainly external (accretion disc and secondary reemission processes), and the scattering occurs in the Thomson regime,  the characteristic velocity depends only on the angular distribution of photons and not on the plasma particle distribution, as explained in Section \ref{sec:gamma_b}. We note however that the situation may be different for type I objects (FR I and BL lacs) where the radiation field is dominated by the jet itself; in this case , there is no simple calculation of the equilibrium velocity. In the following, we only consider FSRQ with an intense external photon field. This does not mean that the two-flow model is invalid for type I objects, but only that the velocity of the pair spine is not easily computed for these objects. In the same way, the application to other kinds of objects such as microquasars and GRBs is probably different given the very different physical conditions of the photon source.

\section{Numerical modelling \label{sec:num_model}}

The numerical model developed here is based on the one described in \cite{Boutelier:2008bga}. The jet is stratified along the axis coordinate $z$ (see Fig. \ref{fig:two-flow}). The jet axis is viewed at an angle $\theta_i$ with the line of sight, with $\mu_i = \cos \theta_i$. Numerically, we define a slicing with an adaptive grid step as described below and physical conditions are computed at each slice, starting from initial conditions given at the basis of the jet.
Each slice then acts as a one-zone and emits synchrotron radiation, synchrotron self-Compton (SSC), and external Compton (EC) radiation, computed as in a one-zone approximation with a spherical blob of radius $R(z)$.
The photon field from external sources (the accretion disc, the dusty torus, and the broad line region) is computed all along the jet. This determines the external inverse-Compton emission as well as the induced Compton rocket force on the plasma as described below. The opacity to high-energy photons inside and outside the jet is computed numerically, and is used to compute a pair creation term
at each step.

\subsection{Geometry of the jet \label{sec:jet_geometry}}

To compute the total jet emission, one needs to know the physical conditions all along the jet. Some of these conditions are derived from the computation of the spatial evolution equation along $z$ (see Sect. \ref{sec:distr_evol}) and only three physical parameters, that is, the inner radius of the jet $R(z)$ (acting as an outer radius for the pair beam), the magnetic field $B(z),$ and the heating term $Q(z)$, are imposed on a global scale. Here we assume their  evolution to be described by power-laws:

\begin{equation}
\label{eq:Rz}
        R(z) = R_0 \left [ \frac{z}{Z_0} + \left(  \frac{R_i}{R_0}\right)^{1/\omega} \right ]^\omega \qquad \text{with} \quad \omega < 2
.\end{equation}

\corr{This law describes a paraboloid with a radius close to $R_{0}$ at a distance $Z_{0}$ from the black-hole. 
The constant $(R_i/R_0)^{1/\omega}$ allows to avoid divergence issues at $z=0$ by setting a minimal jet radius $R(z=0) = R_i$ corresponding to the disc inner radius. The starting and ending altitudes of the jet are free parameters.
% The parabola is displaced by the constant $R_i$\footnote{$R_{i}$ is set at the gravitational radius $R_{g} = GM_{bh}/c^{2}$} to avoid divergence issues when $z$ tends to 0.
}
The index $\omega$ defines the jet opening. One must have $\omega < 1$ to keep the jet collimated.

The magnetic field is supposed to be homogeneous and isotropic at every altitude $z$ in the plasma rest frame. Its evolution is described by:
\begin{equation}
\label{eq:Bz}
                B(z) = B_0 \left( \frac{R(z)}{R_0}\right)^{-\lambda} \qquad \text{with} \quad 1< \lambda < 2
.\end{equation}

The index $\lambda$ gives the structure of the magnetic field and is confined by two extremes:
\begin{itemize}
\item
$\lambda = 1$ corresponds to a pure toroidal magnetic field as the conservation of the magnetic field circulation gives $B \sim 1/R.$
\item
$\lambda = 2$ corresponds to a pure poloidal magnetic field  as the conservation of the magnetic flux in this case gives $B \sim 1/R^{2}$.
\end{itemize}

The particle acceleration is a central part of the two-flow model as it is assumed that the particles are continuously heated by the outer MHD structure (see Sect. \ref{sec:twoflow}) which acts as an energy reservoir, compensating for radiation losses. Due to the lack of a precise expression for the acceleration rate per particle, $Q_{acc}$, we use the following expression:

\begin{equation}
\label{eq:Qz}
                Q_{acc}(z) = Q_0 \left [ \frac{z}{Z_0} + \left(\frac{R_i}{R_0} \right)^{1/\omega}\right ]^{-\zeta} \exp \left( -\frac{z}{Z_c} \right)
.\end{equation}

The particle acceleration decreases as a power-law of index $\zeta$ up to an altitude $z=Z_{c}$ where it drops exponentially. This altitude $Z_{c}$ physically corresponds to the end of the particle acceleration (through turbulence) in the jet.
Because of this exponential cut-off, whatever the index $\zeta$ is, the total amount of energy provided to accelerated particles remains finite. However, as $Z_{c}$ could be as large as desired (even as large as the jet), we consider $\zeta > 1$ to be physically more satisfactory. This way, even an integration to infinity of $Q_{acc}$ would converge. Similarly to the jet radius expression (equation \ref{eq:Rz}), the constant $R_i/R_0$ avoids numerical issues for very small $z$.

The jet is then sliced along $z$. As we see below, the particle density can be subject to abrupt changes in case of \corr{short and intense events of pair creation which is a non-linear process}. It is therefore essential to have an adaptive slicing as the physical conditions are computed in the jet. \corr{The condition have been chosen to ensure variation rates of the particle mean energy and of the particle flux of less than $1\permil$ between two computation steps}.

Frequencies follow a logarithmic sampling between $\nu_{min}$ and $\nu_{max}$ in the observer frame. The sampling is therefore different at each altitude in the jet depending on the Lorentz factor. This ensures that local emissivities are computed for the same sampling (that of the observer) when transferred to the observer frame.

\subsection{Geometry of the external sources of photons \label{sec:ext_sources}}

There are several possible sources of soft photons in an AGN but three have an actual influence on the external Compton emission and are taken into account in our model: the accretion disc, the dusty torus, and the broad line region (BLR). In order to correctly compute the corresponding inverse-Compton radiation, the anisotropy of the sources is taken into account as detailed below. 

\subsubsection{The accretion disc \label{sec:accrection_disc}}

The geometry of the disc \corr{(see Fig. \ref{fig:sketch_disc})} is described by its internal radius $R_{in}$ and its external radius $R_{out}$. It is then sliced along its radius $r$ (with a logarithmic discretization) and its azimuthal angle $\varphi$ (with a linear discretization). Therefore, each slice has a surface $\displaystyle \d S = \d \varphi \left(r\d r + \frac{\d r^2}{2}\right)$.
From the jet axis at an altitude $z$, each slice is seen under a solid angle 
\begin{equation}
\d \Omega = \d S \frac{z}{\left( r^2 + z^2\right)^{3/2}}
,\end{equation}
\corr{at an angle $\theta_s = \arccos(z/\sqrt{r^2 + z^2})$ with the $z$ axis.}
We consider here a standard accretion disc. \corr{Then the radial distribution of the temperature is given by $T_{disc}(r)$ \citep{1976MNRAS.175..613S}:
\begin{equation}
\label{eq:T_disc}
T_{disc}(r) = \left[ \frac{3 G M \dot{M}} { 8 \pi \sigma} \frac{1}{r^3} \left(1 - \sqrt{\frac{R_{isco}}{r}} \right)   \right] ^{1/4}
,\end{equation}
with $G$ the gravitational constant, $\sigma$ the Stefan-Boltzmann constant, $M$ the black-hole mass, $\dot{M}$ the accretion rate, and $\displaystyle R_{isco} = \frac{6MG}{c^2}$ the innermost stable circular orbit. Its emissivity is equal to one if not specified otherwise.

The luminosity of one face of the disc is then given by
\begin{equation}
\label{eq:disc_luminosity}
L_{disc} = \int_{R_{in}}^{R_{out}} \sigma T^4_{disc}(r) 2 \pi r \dd r
,\end{equation}
which converges to $\displaystyle L_{disc} = \frac{\dot{M}c^2}{24}$ for $R_{in} = R_{isco}$ and $R_{out} \gg R_{in}$.}

\begin{figure}[h]
\centering
        \includegraphics[width=\hsize]{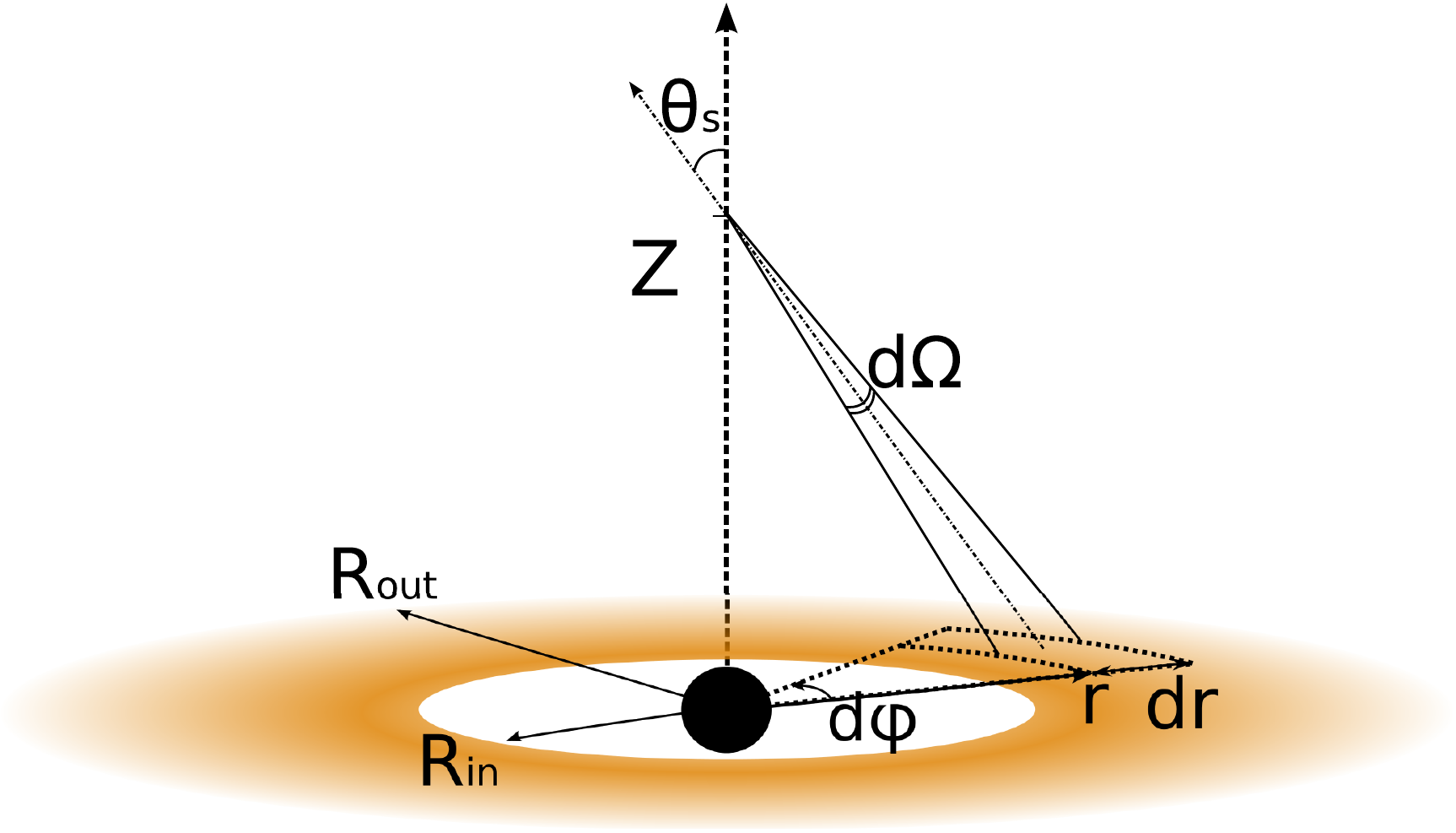}
        \caption{\label{fig:sketch_disc} Sketch of the disc radial and azimuthal splitting. A slice at $(r,\varphi) \in \left( [R_{in},R_{out}],[0,2\pi]\right)$ is seen under a solid angle $\d \Omega$ from an altitude $z$ in the jet.}
\end{figure}

\subsubsection{The dusty torus}

The dusty torus (see Fig. \ref{fig:torus}) is assumed to be in radiative equilibrium with the luminosity received from the accretion disc. The torus is sliced according to $\theta_t \in \left[\theta_{t_{min}},\theta_{t_{max}}\right]$ and $\varphi \in [0,2\pi]$ . Each slice is illuminated by the disk and is in radiative equilibrium. It emits as a grey-body with a temperature $T_t(\theta_t,\varphi)$ and an emissivity $\varepsilon(\theta_t, \varphi)$. As most of the disc luminosity comes from its inner parts and considering $R_{in} \ll (D_t - R_t)$, the disc appears point-like from a torus surface point. Therefore, each slice of the torus of surface $\d S_t$ and at an angle $\theta_t$ from the $X$ axis verifies the relation:
\begin{equation}
\label{eq:torus_eq}
I_d \: S_d \: \d \Omega_d \: \cos \omega_d = \varepsilon(\theta_t) \sigma T_{torus}^4(\theta_t) \d S_t (\theta_t)
,\end{equation}
with $\omega_d$ the angle between the Z-axis and the emission direction from the disc: $\displaystyle \cos \omega_d = a \sin \theta_t / \sqrt{1 - 2 a \cos \theta_t + a^2}$ and $a = R_t / D_t$.

\begin{figure}[ht]
        \includegraphics[width=\hsize]{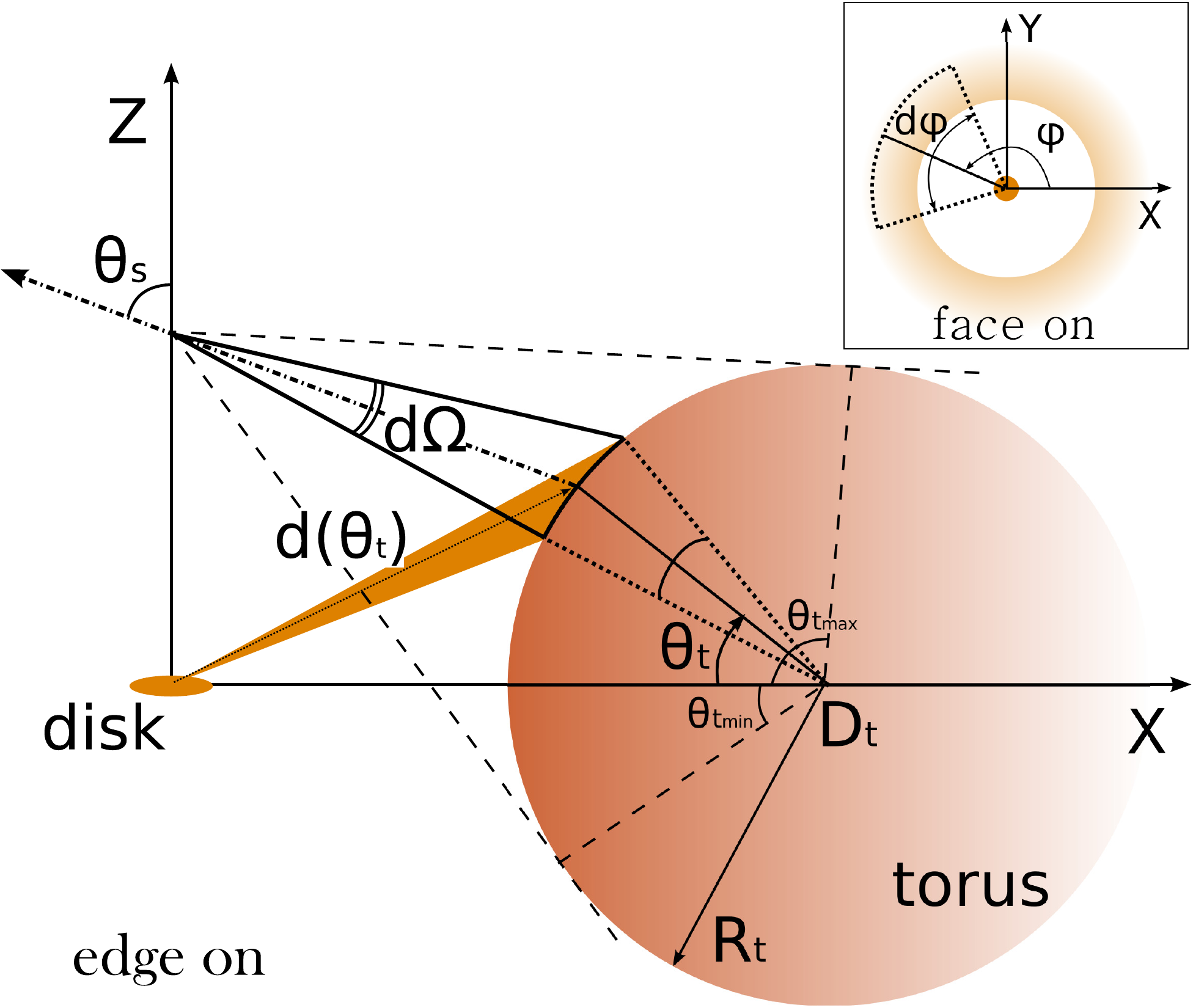}
\caption{\label{fig:torus}Dusty torus sliced as described in the text. Each slice is seen from an altitude $z$ in the jet under a solid angle $\d \Omega$ . }
\end{figure}

\corr{The user can then either fix a constant temperature or a constant emissivity and compute the other parameters as a function of $\theta_t$ to preserve the radiative equilibrium. If not specified otherwise, $\varepsilon(\theta_t) = 1$ and the temperature is kept free.}

\subsubsection{The broad line region}

The broad line region is modelled by an isotropic, optically and geometrically thin spherical shell of clouds situated at a distance $R_{blr}$ from the central black-hole. Like other sources, it is sliced angularly into different parts in order to perform the numerical integration. We choose a linear discretization along $\omega \in \left[\omega_{max},\omega_{min} \right]$ and along $\varphi \in [0:2\pi]$.

\begin{figure}[ht]
        \includegraphics[width=\hsize]{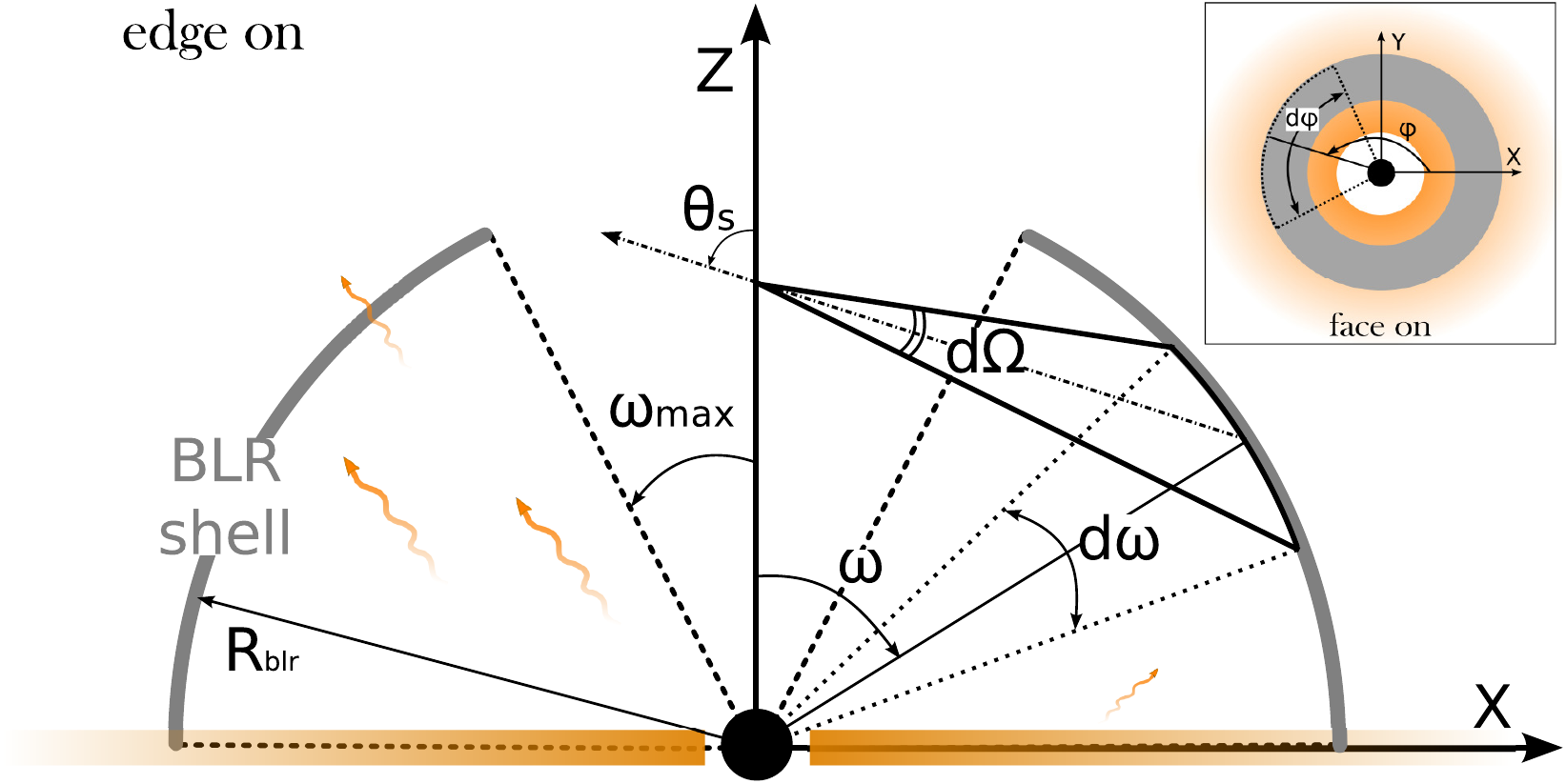}
\caption{\label{fig:blr}The BLR, an optically and geometrically thin shell of clouds seen under a solid angle $\d \Omega$ from an altitude $z$ in the jet. The BLR is sliced according to $\omega \in \left[\omega_{max},\pi/2\right]$ and $\varphi \in [0,2\pi]$. The BLR absorbs and re-emits part of the disk luminosity.}
\end{figure}

Observed BLRs display a complex emission with a continuum and broad absorption lines but \cite{Tavecchio:2008fu} showed that modelling the spectrum of the BLR with a grey-body spectrum at $T=10^5 K$ provides a good approximation to the resulting inverse-Compton spectrum. \corr{We followed this idea using a temperature of $T_{blr} = 10^5 K$ and an overall luminosity $L_{blr}$, which is a fraction $\alpha_{blr}$ of the disc luminosity: $\displaystyle L_{blr} = \alpha_{blr} L_{disc} = \sigma T_{blr}^4 R^2 \int_{0}^{2\pi}\dd \phi \int_{\omega_{max}}^{\pi/2}\sin(\omega) \dd \omega$.}\\
Emissivity of the BLR is then given by:
\begin{equation}
        \varepsilon_{blr} = \frac{\alpha_{blr} L_{disc}}{\sigma T^4_{blr} 2\pi R^2_{blr}\cos \omega_{max}}
        \label{eq:iso_emissivity}
.\end{equation}

\subsection{Emission processes}

As the spine is supposed to be filled by electrons/positrons, only leptonic processes need to be considered here: synchrotron and inverse-Compton radiation are computed all along the jet. The radiation is computed in the plasma rest frame and then boosted by the local Doppler factor $\delta(z)=(\Gamma_b(z)(1-\beta_b(z)\cos i_{obs})^{-1}$ into the observer's rest frame - with $\beta_b = \sqrt[]{1 - 1/\Gamma_b^2}$ and $i_{obs}$ the observer angle relative to the jet axis as defined in Fig. \ref{fig:absorption_sketch}.\\

In the two-flow paradigm, particles are supposed to be accelerated by a  second-order Fermi process due to turbulence inside the jet. Thus a Fokker-Planck equation governs the evolution of the particle distribution which evolves by diffusive acceleration and radiative losses. \citep{Schlickeiser:1984ua} showed that generic solutions of this equation were pile-up (or quasi-Maxwellian) distributions. Most of the particles are then concentrated around some characteristic energy $\bar{\gamma}$ where the acceleration and cooling time are similar. We adopt the following simplified form of such a distribution:
\begin{equation}
        n_{e}(\gamma, z) = n_{0}\left(z\right) \frac{\gamma^{2}}{2 \bar{\gamma}^{3}(z)} \exp \left(-\gamma/\bar{\gamma}(z)\right)
\label{eq:pileup}
,\end{equation}

where $\displaystyle n_{0}(z) = \int n_{e}(\gamma) \dd \gamma$.\\

\corr{As this acceleration mechanism does not produce power laws for the particles energy distributions (contrary to the first-order Fermi process at work in shocks), the reproduction of large-scale power laws in the spectra of the sources is not straight forward. Power-law shapes can however be reproduced by a summation of a number of pile-up (from different emission zones) with different parameters.
% Note that in this mechanism, contrary to the common assumption power-law are not produced locally but will be the result of power-law distributions of parameters along the jet.
However, the summation of the emission coming from each slice of the jet can be demanding in terms of computing time.} A\ certain level of approximation is thus required to achieve computation of the model in a reasonable amount of time. The following subsections present the calculation of each type of radiation for a pile-up distribution.

\subsubsection{Synchrotron radiation}

%\citep{Sauge:2004tc} showed that for a pile-up distribution, the synchrotron spectrum follows three regimes given by:
%
%\begin{itemize}
%\item
%$\nu < \nu_{abs}$: the medium is optically thick and the spectrum follows a power-law
%\begin{equation}
%       F_\nu = \frac{8\pi m_e}{R} \bar{\gamma} \nu^2
%\end{equation}
%
%\item
%$\nu_{abs} < \nu < \bar{\nu}_c$: the medium is optically thin and the spectrum is described by a power-law of index $-1/3$
%\begin{equation}
%       F_\nu = \frac{16 \pi^2 2^{2/3}}{27} \frac{e^2 N_e \bar{\gamma}}{c} \left( \frac{\nu}{\bar{\nu}_c} \right)^{-1/3}
%\end{equation}
%
%\item
%$\nu > \bar{\nu}_c$: the emission drops exponentially
%\end{itemize}
%
%with $\displaystyle \bar{\nu}_c = \frac{3}{4 \pi} \frac{e B \bar{\gamma}^2}{m_e c}$ the critical synchrotron frequency.
%
%The complete calculation has been done in \citep{Sauge:2004tc} and recalled in the appendix.
%
%\corr{DEV - calculation in Annexe}
%
%The synchrotron emissivity is given by (\corr{source - SaugeThesis to cite but give all calculation !}):
%\begin{equation}
%       j_\nu = \frac{1}{2} \frac{\sqrt{3}e^2 B \sin \theta}{4 \pi m_e c^2} \int \d \gamma n(\gamma) F_{syn} \left( \frac{\nu}{\sin \theta \nu_c}\right)
%\label{eq:syn_em}
%\end{equation}
%
%
%
%For a pile-up distribution, one obtains 
%
%\begin{equation}
%       I(y) = \frac{8 \pi}{9 \sqrt{3}} \left( \frac{y}{2}\right)^{-2/3}
%\end{equation}
%
%$------------------------------------$\\

Using the expression of the synchrotron-emitted power per unit of frequency and solid angle (\citep{BLUMENTHAL:1970gb}), \citep{Sauge:2004tc} showed that the synchrotron emissivity of a pile-up distribution can be written as:
\begin{equation}
        J_{syn} = \frac{\sqrt{3}}{16 \pi} \frac{e^3 B}{m_e c^2} N_e y \: \Lambda(y)
    \label{eq:Jsyn}
,\end{equation}
% with $\displaystyle  \bar{\nu}_c = \frac{3}{4 \pi} \frac{e B \bar{\gamma}^2}{m_e c}\: $,  $\displaystyle y = \nu/\bar{\nu}_c$
with $\displaystyle  \eps_c = \frac{3}{4 \pi} \frac{h e B \bar{\gamma}^2}{m_e^2 c^3}\: $,  $\displaystyle y = \eps/\eps_c$
and
\begin{equation}
        \Lambda (y) = \frac{1}{y} \int_0^\infty x^2 \exp(-x) F_{syn}\left(\frac{y}{x^2}\right) \d x
.\end{equation}

To fasten numerical computation, analytical approximations of the function $\Lambda(y)$ can be done in different regimes. These approximations are presented in the appendix (equations \ref{eq:lambda1}, \ref{eq:lambda2} and \ref{eq:lambda3}).

% Finally, one obtains the specific intensity of a portion of jet of radius $R$ in the optically thin regime:
% % \begin{equation}
% % \label{eq:sync_thin}
% %     I_\nu^{thin} (\nu) = R j_\nu = \frac{\sqrt{3}}{16 \pi} \frac{e^3 B R}{m_e c^2} N_e y \: \Lambda(y)
% % \end{equation}
% \begin{equation}
% \label{eq:sync_thin}
%       I_\eps^{thin} (\eps) = R j_\eps = \frac{\sqrt{3}}{16 \pi} \frac{e^3 B R}{m_e c^2} N_e y \: \Lambda(y)
% \end{equation}

Finally, one obtains the synchrotron spectrum in the optically thin regime $(\eps > \eps_{abs})$:
% \begin{equation}
% \label{eq:sync_thin}
%       I_\nu^{thin} (\nu) = R j_\nu = \frac{\sqrt{3}}{16 \pi} \frac{e^3 B R}{m_e c^2} N_e y \: \Lambda(y)
% \end{equation}
\begin{equation}
\label{eq:sync_thin}
\frac{\d n_s^{thin}(\eps)}{\d \eps \d t} = \frac{4 \pi}{\eps_1 m_e c^2} J_{syn}=
\frac{\sqrt{3}}{4} \frac{e^3 B}{m_e^2 c^4} N_e \frac{1}{\eps_c} \: \Lambda(y)
.\end{equation}

The transition between the optically thin and the optically thick regime happens at the frequency $\eps_{abs}$ defined by 
\begin{equation}
\label{eq:tau_syn_def}
\tau_\eps(\eps_{abs}) \approx \frac{I_\eps^{thin}(\eps_{abs}) h^2}{2 \bar{\gamma} m_e^3 c^4 \eps_{abs}^2} = 1
.\end{equation}

For a pile-up distribution of particles, the distribution temperature $k_B T_e = \langle \gamma \rangle m_e c^2 /3 = \bar{\gamma} m_e c^2$ does not depend on $\eps$. In this case, the optically thick regime at low frequency $(\eps < \eps_{abs})$ is described by the Rayleigh-Jeans law:
% \begin{equation}
% \label{eq:sync_thick}
% I_\nu^{thick} (\nu) = 2 \bar{\gamma} m_e \nu^2
% OR I_\eps^{thick} (\eps) = 2 \bar{\gamma} \frac{m_e^3 c^4}{h^2} \eps^2
% \end{equation}
\begin{equation}
\label{eq:sync_thick}
\frac{\d n_s^{thick}(\eps)}{\d \eps \d t} = \frac{8 \pi}{R} \frac{m_e^2 c^2}{h^2} \bar{\gamma} \eps
.\end{equation}

Finally the synchrotron spectrum resulting from a pile-up distribution has three main parts:
\begin{itemize}
\item $\eps < \eps_{abs}$ is the optically thick part of the spectrum described by a power-law of index 1.
\item $\eps_{abs} < \eps < \eps_c$ is the optically thin part of the spectrum described by a power-law of index -2/3.
\item $\eps_{c} < \eps $ is where the spectrum falls exponentially.
\end{itemize}

%For $\displaytyle \nu < \nu_{abs}$ (defined as ), one enters the optically thick regimes. 

\subsubsection{Synchrotron self-Compton radiation}

We assume the synchrotron self-Compton (SSC) emission to be co-spatial with the synchrotron emission, and therefore neglect the synchrotron emission coming from other parts of the jet. We treat the Thomson and the Klein-Nishina (KN) regimes separately, following the analytical approximations proposed by \citep{Sauge:2004tc}. The results of these approximations are recalled here while the details are given in the appendix. We first consider the distinction between the Thomson and the KN regimes relative to the synchrotron peak $\eps_c$.\\

If $\eps_c \ll 1/\eps_1$, all synchrotron photons producing SSC photons of energy $\eps_1$ are scattered in the Thomson regime. In this case, one can show that the SSC photon production rate is given by (see appendix \ref{sec:appendix_ssc}):
\begin{equation}
\label{eq:SSC_rate_th}
\frac{\d n_{ssc}^{th}}{\d \eps_1 \d t} = \frac{3 \sigma_{th}}{4} \frac{n_0}{2\bar{\gamma}} \tilde{G}\left(\frac{\eps_1}{\bar{\gamma}^2} \right)
,\end{equation}
with
\begin{equation}
\label{eq:ssc_G}
\begin{aligned}
\tilde{G}(x) & = \int \frac{\d \eps}{\eps} \frac{\d n_{ph}}{\d \eps} \tilde{g}\left( \frac{x}{4\eps} \right) \\
\tilde{g}(x) & = \frac{2}{3} e^{-\sqrt{x}}\left( 1 - \sqrt{x} + xe^{\sqrt{x}} E_i(\sqrt{x}) \right)
\end{aligned}
,\end{equation}
with $E_i(x) = \int_x^\infty \d t \, e^{-t}/t $ being the exponential integral function. Interestingly, $\tilde{G}$ is a function of a single variable and can therefore be tabulated to speed up calculation in the Thomson regime.\\

For $\eps_c > 1/\eps_1$, we must take into account KN corrections. However, synchrotron photons verifying $\eps < 1 / \eps_1$ are still in the Thomson regime and photons verifying $ \eps > 1/\eps_1 $ are in the KN regime. In order to include KN corrections only when necessary, the emissivity in this regime is divided into two contributions, Thomson and Klein-Nishina, respectively given by $J_{ssc}^{th}$ and $J_{ssc}^{kn}$:
\begin{equation}
\label{eq:ssc_kn_th}
J_{ssc}^{th}(\eps_1) =  \eps_1^{-s} \bar{\gamma}^{2s+1} \left(\Gamma(2s+1, u_{min}) - \Gamma(2s+1, u_{max}) \right)
,\end{equation}
\corr{with $\Gamma$ being the incomplete gamma function and}
with $\displaystyle u_{min} = \sqrt{\frac{\eps_1}{\eps_{max}\bar{\gamma}^2}}$ and $\displaystyle u_{max} = \sqrt{\frac{\eps_1}{\eps_{min}\bar{\gamma}^2}}, $

% The Klein-Nishina contribution is given by:
\begin{equation}
\label{eq:ssc_kn_kn}
\begin{aligned}
J_{scc}^{kn}(\eps_1) & = \frac{3}{8} \eps_1 \exp\left(-\frac{\eps_1}{\bar{\gamma}}\right) \\
& \times \left\{ \left(\ln(2\eps_1) +\frac{1}{2} \right) K_1^{(s)}(1/\eps_1, \eps_0) +  K_2^{(s)}(1/\eps_1, \eps_0) \right\}. 
%\left{ (\ln(2\eps_1) +\frac{1}{2} ) K_1^{(s)}(1/\eps_1, \eps_0) +  K_2^{(s)}(1/\eps_1, \eps_0)\right}
\end{aligned}
\end{equation}

The SSC photon production rate in the Klein-Nishina regime is then given by
\begin{equation}
\label{eq:ssc_rate_kn}
\frac{\d n_{ssc}^{kn} (\eps_1)}{\d \eps_1 \d t} = n_1 n_0 c \sigma_{Th} \left( J_{scc}^{th}(\eps_1) + J_{ssc}^{kn}(\eps_1) \right)
.\end{equation}

The continuity between the two regimes given by Eqs. \ref{eq:SSC_rate_th} and \ref{eq:ssc_rate_kn} is assured by \corr{an interpolation formula} that gives the complete expression of the SSC radiation:
%n this section, we consider the dimensionless energy (as introduced before) $\displaystyle \eps = h\nu/m_e c^2$

\begin{equation}
\label{eq:ssc_continuity}
\cfrac{\d n_{ssc}(\eps_1)}{\d \eps_1 \d t} = \cfrac{ \cfrac{\d n_{ssc}^{th}(\eps_1)}{\d \eps_1 \d t} + x^n \cfrac{\d n_{ssc}^{kn}(\eps_1)}{\d \eps_1 \d t} }{1 + x^n}
\qquad \text{with} \quad x = \eps_{1}\eps_c
.\end{equation}
\corr{We used some examples to verify that the value $n = 6$ gives a correct approximation of the full cross section. However the final results are relatively insensitive to the choice of $n$ since the various contributions are smoothed by the spatial convolution of the different parts of the jet.} 

\subsubsection{External Compton radiation}

The calculation of the inverse-Compton emission on a thermal distribution of photons (a.k.a. external Compton) is the most demanding in terms of computation time since it requires an integration over the energy and spatial distributions of the incoming photons which are not produced locally. We note that the anisotropy of the incoming radiation is important and has to be properly taken into account for the computation of the emissivity and the bulk Lorentz factor of the plasma through the Compton Rocket effect (see section \ref{sec:gamma_b}).

Since all our external emission sources (disk, BLR and torus) can be approximated by a blackbody energy distribution(or the sum of several blackbodies),  some approximations are possible and have been proposed by \cite{Dubus:2010ez}, \cite{2013arXiv1310.7971K}, and \cite{Zdziarski:2013ed} (hereafter ZP13). The method we developed and used in this paper is less precise than the one proposed by ZP13 but is at least twice  faster; and up to ten times faster in some cases. It  approximates the Planck's law for black-bodies emission by a Wien-like spectrum. Details of the calculation and comparison with ZP13 are done in Appendix B.\\

The number of inverse-Compton photons per energy unit and time unit produced by the scattering of a thermal photon field on a single particle of energy $\displaystyle \gamma m_e c^2$ is then given by:
\begin{equation}
\label{eq:ic_wien_final}
   \begin{aligned}
        \frac{\d N_{ec} (\eps_1)}{\d t \d \eps_1} = & \frac{m^2_e c^4 r^2_e A \epsilon' \pi}{h^3 \gamma^4 (1-\beta\cos\theta_0)} \frac{1}{(1-x)} \left\lbrace \left( \frac{2}{\mathcal{H}} + 2 + x\bar{\epsilon}'\right) e^{-\mathcal{H}/2} \right.\\
          & \left. - \frac{}{} \left(2+\mathcal{H}\right) \left(E_1(\mathcal{H}/2)-E_1(2\gamma^2 \mathcal{H})\right) \right\rbrace
   \end{aligned}
,\end{equation}
with $\di x = \frac{\eps_1}{\gamma}$, $\theta_0$ being the angle between the incoming photon and the particle direction of motion,
$\di \mathcal{H}=\frac{x}{\bar{\epsilon}'(1-x)}  $
, and $\di E_1(x) = \int_x^\infty \frac{\exp(-t)}{t} \d t$ being the exponential integral.
\corr{This relation takes into account the anisotropy of the emission through the angle $\theta_0$ and the full Klein-Nishina regime.}\\

Equation (\ref{eq:ic_wien_final}) needs to be integrated over the pile-up distribution to get the complete spectrum. This can be long in the general case. However, \corr{when applicable (we chose the conservative limit of $\epsilon' < 0.01$)}, which greatly simplifies in the Thomson regime (see Appendix \ref{sec:appendix_ec}) and one obtains:
\begin{equation}
\label{eq:ec_pileup_thomson}
\frac{\d n_{ec}^{Th}}{\d t \d \eps_{1}} = \frac{\pi m_{e}^{2}c^{4}r_{e}^{2}A\eps_{1} }{h^{3}}  \frac{N_e} {2\bar{\gamma}^4 (1-\mu_0) } \:  \chi(s)
,\end{equation}
with 
\begin{equation}
\label{eq:chi}
\begin{aligned}
\chi(s) = \int_0^\infty & \frac{e^{-u}}{u^{2}} 
 \left\lbrace \left( \frac{2u^2}{s} + 2 \right) \exp\left(-\frac{s}{2u^2}\right) \right. \\
         & \left. - \left( 2+ \frac{s}{u^2} \right) E_1 \left(\frac{s}{2u^2}\right) \right\rbrace
\d u 
\end{aligned}
,\end{equation}
with $\chi(s)$ being a single variable function. As such, it can be computed once and then tabulated and interpolated over when required. As a result, \corr{when the emission occurs in the Thomson regime,} the computation of the inverse-Compton spectra from the scattering of a thermal soft photon field on a pile-up distribution of electrons can be done much faster than usual with complete numerical integration.

\subsection{Photon-photon absorption in the jet and induced pair creation}

High-energy photons produced by inverse Compton will locally interact with low-energy photons produced by the synchrotron process (external radiations can be locally neglected). This photon-photon interaction induces an absorption that needs to be taken into account to compute the emitted flux and the pair creation process loading the jet with leptons.

\subsubsection{Escape probability in the jet \label{sec:abs_jet}}

For a photon of dimensionless energy, $\displaystyle \eps_1= h \nu_1/m_e c^2$ in a photon field of photon density per solid angle, and per dimensionless energy, $n_{ph}(\eps_2,\Omega)$, the probability to interact with a soft photon of energy $\eps_2$ on a length $\d l$ is given by:
\begin{equation}
\label{eq:dTaugg}
        \frac{d}{\d l} \tau_{\gamma\gamma}(\eps_1) = \int \frac{1-\mu}{2} \, \sigma(\beta) \, n_{ph}(\eps_2, \Omega) \: \d \eps_2 \: \d \Omega
,\end{equation}
with $\sigma(\beta)$ being the interaction cross-section given by \citep{Gould:1967bj}:
\begin{equation}
\label{eq:sigma_gould}
\begin{aligned}
& \sigma(\beta) = \frac{3 \sigma_{Th}}{16} \left( 1 - \beta^2 \right)
\left[ (3 - \beta^2) \ln \left( \frac{1 + \beta}{1-\beta}\right) - 2 \beta (2 - \beta^2)  \right] \\
& \beta(\eps_1, \eps_2, \mu) = \sqrt{1 - \frac{2}{\eps_1 \eps_2 (1- \mu)}}
\end{aligned}
,\end{equation}
with $\mu$ being the cosine of the incident angle between the two interacting photons in their reference frame.\\

If we make the approximation that the synchrotron emission is locally isotropic in the plasma rest frame, the optical depth per unit length $\d l$ simplifies and is then given by \citep{Coppi:1990tn}:
\begin{equation}
\frac{\d}{\d l} \tau_{\gamma\gamma}^{jet}(\eps_{1}) = \frac{1}{c} \int \d \eps_2 n_{ph}(\eps_2) R_{pp}(\eps_1 \eps_2)
,\end{equation}
with $R_{pp}$ the angle-averaged pair production rate $(cm^3.s^{-1})$.

Analytical approximations of $R_{pp}$ have been proposed by \cite{Coppi:1990tn} and \cite{Sauge:2004ep} but they still require a timeconsuming integration over all energies. In order to simplify numerical calculations, as the function $R_{pp}$ is peaked around its maximum $R_{pp}^{max}$ occurring at $x = x_{max}$, we make the approximation:
\begin{equation}
R_{pp}(x) = R_{pp}^{max} \, \delta(x - x_{max})
\end{equation}
with
\begin{equation}
\begin{aligned}
R_{pp}^{max} & = 0.283 \, \frac{3}{4} c \sigma_{Th} \\
x_{max} & = 3.5563 ,
\end{aligned}
\end{equation}
with $\sigma_{Th}$ being the Thomson cross-section.\\

The optical depth in the jet finally simplifies to
\begin{equation}
\label{eq:dtau_approx}
        \frac{\d}{\d l} \tau_{\gamma\gamma}^{jet}(\eps_{1}) = \frac{1}{c} \frac{R_{pp}^{max}}{\eps_{1}} n_{ph} \left(     \frac{x_{max}}{\eps_{1}}         \right)
.\end{equation}

Assuming that the absorption coefficient is constant at a given altitude $z$ in the jet of radius $R(z)$, the opacity can be calculated as
\begin{equation}
\label{eq:tau_jet}
\tau_{jet}(\eps_1, z) = R(z) \frac{\d \tau^{jet}_{\gamma \gamma}}{\d l} \left(\eps_{1},z \right)
.\end{equation}

Solving the transfer equation in the plane-parallel approximation for photons absorbed in-situ gives their escape probability $\displaystyle \Pp_{esc}^{jet}$:
\begin{equation}
        \label{eq:escape_jet}
        \Pp_{esc}^{jet}(\eps_1, z) = \left( \frac{1-\exp \left(-\tau_{jet}(\eps_1, z) \right)}{\tau_{jet}(\eps_1, z)}      \right)
.\end{equation}
This escape probability is useful for computing the pair production rate inside the jet. However, another absorption factor must be considered in the vicinity of the jet as discussed in Sect. \ref{sec:abs_out}, since the photon density does not vanish abruptly outside the jet.

\subsubsection{Pair creation}

The pair production is a direct consequence of the $\gamma-\gamma$ absorption in the jet. A photon of dimensionless high-energy $\eps > 1$ interacts preferentially with a soft photon of energy $\eps \approx 1/\eps$ to form a leptonic pair $e^+/e^-$. Both created particles have the same energy $\gamma m_e c^2$ (with $m_e$ the electron mass) and one can write the energy conservation as $\eps + 1/\eps \approx \eps = 2\gamma$.

Therefore, at a given altitude of $z$, the pair production rate $\dot{n}_{prod}$ is given by
\begin{equation}
        \label{eq:nprod}
        \dot{n}_{prod}(z)  =   2 \int \d \eps_{1} \left. \frac{\d n(\eps_{1})}{\d \eps_{1} \d t}\right|_z (1-\Pp_{esc}^{jet}(\eps_1, z))
.\end{equation}

This pair creation is then taken into account in the evolution of the particle density. The created particles are not supposed to cool freely but they are rather constantly reaccelerated by the turbulence. We assume that the acceleration process is fast enough to maintain a pile-up distribution without considering the perturbation to the energy distribution introduced by the pair creation.

\subsection{\label{sec:distr_evol}Evolution of the particle distribution}

The pile-up distribution has two parameters, the \emph{particle density} $n_{0}(z)$ and the \emph{particles characteristic energy} $\bar{\gamma}(z) m_{e}c^{2}$ (the particles mean energy being given by $3 \bar{\gamma} m_{e} c^{2}$). Both parameters are not imposed but computed consistently all along the jet as detailed below.

\subsubsection{Evolution of the particles' characteristic energy}

The particles' characteristic energy $\bar{\gamma}(z) m_{e} c^{2}$ results from the balance between heating from the turbulence and radiative cooling. The energy equilibrium for each particle in the comoving frame can be written as follows.
\begin{equation}
\label{eq:energy_eq_particle}
\frac{\partial}{\partial t} \left( \bar{\gamma} m_e c^2\right) = \frac{1}{\Gamma_b} \left( Q_{acc} m_e c^2 - P_{cool}\right)
,\end{equation}

with the acceleration parameter $Q_{acc}$ $(s^{-1})$ (see Eq. \ref{eq:Qz}) and $P_{cool}$ being the emitted power derived from synchrotron emission, SSC, and EC emission. We consider here that the cooling is efficient and occurs mainly in the Thomson regime, neglecting the cooling in the Klein-Nishina regime. In the case of isotropic distribution of photons, \cite{1986rpa..book.....R} showed that the emitted power at the characteristic particles energy writes $\displaystyle P = \frac{4}{3} c \sigma_{T} U \left({\bar{\gamma}}^{2}-1\right)$ with U being the total energy density of the photon field. One must compute the contribution of each energy density corresponding to each emission process: $\displaystyle U = U_B + U_{syn} + U_{ext}$.

% \corr{on ne considere pas dans le code une densite effective limitee a $\epsilon < 1/\bar{\gamma}$ pour rester dans le regime Thomson ?}

For the synchrotron emission, we consider an isotropic magnetic field. \cite{1986rpa..book.....R} gives
\begin{equation}
        U_{B} = \frac{B^{2}}{8\pi}
\label{eq:ub}
.\end{equation}

% For the synchrotron and the SSC emission, one can consider the emission and the energy density isotropic in the comoving frame and therefore we have:

The power emitted through SSC is computed considering the effective energy density of the synchrotron photon field corresponding to the Thomson regime (using a cut-off frequency $\nu_{kn}$). As the synchrotron emission is isotropic and co-spatial with the SSC emission, one obtains
\begin{equation}
        U_{syn} = 4\pi \frac{m_e c}{h} \int_{0}^{\eps_{kn}} I_{\eps}^{syn}(\eps) \dd \eps \qquad \text{with} \quad \eps_{kn} = \frac{1}{\bar{\gamma}}
        \label{eq:usyn}
,\end{equation}
with $ I_{\eps}^{syn}$ being the local synchrotron specific intensity.\\

The external photon field is not isotropic and one should integrate over all directions to derive the power emitted through external inverse-Compton. To ease numerical computation, we assume that the total emitted power at each altitude $z$ can be approximated by the power emitted in the direction perpendicular to the jet axis and integrated over 4$\pi$:  
\begin{equation}
U_{ext} = \frac{3}{4 c \sigma_{T} \left(\bar{\gamma}^{2}-1\right) } \int_0^{\eps_{kn}} \eps \left. \frac{\d N}{\d t \d \eps} \right|_{\mu = 0} \: \d \eps
,\end{equation}

with $\displaystyle \frac{\d N}{\d t \d \eps} \left(t, \eps \right)$ being the emitted external Compton spectrum in the comoving frame.
% We checked that this assumption give estimates of the total emitted power that differs from the exact integration of the angle-dependent power emission by a factor of the order of a few.
% \corr{?a few? ca me semble beaucoup, tu es sur ?}
Finally, we compute the characteristic particle Lorentz factor by
numerically solving the following energy equation in the comoving frame:
\begin{equation}
\label{eq:balance}
\begin{aligned}
        \frac{\partial\bar{\gamma}}{\partial t}(z,t) & =  \delta_{b}(z) \left[Q_{acc}(z) \right.\\
        & - \left. \frac{4}{3} \frac{\sigma_{T}}{m_{e}c} \left(  U_{B} + U_{sync}(\bar{\gamma}) + U_{ext}(\bar{\gamma})  \right)\left( \bar{\gamma}^{2} -1 \right) \right]
\end{aligned}
.\end{equation}

This equation is solved in the stationary regime in each slice of the jet using a Runge-Kutta method of order 4. 

\subsubsection{Evolution of the particles density}

The particle density evolves with the pair production and annihilation. As the particles move forward in the jet, one can apply flux conservation to compute the evolution of the density along the jet.

In the absence of pair production, the particle flux $\displaystyle \Phi_e(z,t) = \int n_{e}(\gamma,z,t) \: \pi R^{2}(z) \: \Gamma_{b}(z) \, \beta_{b}(z) \, c \: \d \gamma$ is conserved.

By generalizing the standard continuity equation, \citep{Boutelier:2009uh} showed that for a stationary jet structure, the flux conservation can be written as follows.
\begin{equation}
\label{eq:flux_conservation}
                D_{\beta_b}\Phi_e = \frac{\partial}{\partial t} \Phi_e + c \beta_b \frac{\partial}{\partial z} \Phi_e = \beta_b S \dot{n}_{prod}
,\end{equation}

with $S(z) = \pi R^{2}(z)$ being the section of the jet and $\dot{n}_{prod}$ the pair production rate given by equation \ref{eq:nprod}.
In the stationary case, one simply solves $\displaystyle \frac{\partial}{\partial z}\Phi_e = \frac{1}{c} S \dot{n}_{prod}$.

As the new particles are created in the jet filled with turbulences, they are accelerated and can emit more radiation, creating more particles. This process is highly non-linear and can lead to explosive events. In the two-flow paradigm, this can lead to fast and powerful flares as the particle density will increase as long as the energy reservoir is not emptied.

\subsection{Evolution of the bulk Lorentz factor of the jet\label{sec:gamma_b}}

\corrtwo{As shown by \citealt{1981ApJ...243L.147O}, the radiative forces act on a hot plasma dynamic through the Compton rocket effect. In the present case, the timescale of the Compton rocket effect is equal to the inverse-Compton scattering time of a photon field of energy density $\displaystyle U_{ph}$ on a particle of energy $\displaystyle \gamma m_e c^2$ and thus can be evaluated as:}
\begin{equation}
t_{IC} = \frac{3 m_e c^2}{4 c \sigma_T \gamma U_{ph}}
.\end{equation}
In the inner regions of powerful AGNs, the photon field energy density at a distance $z$ in the jet can be evaluated as
\begin{equation}
U_{ph} \approx \frac{L_{disc}}{4\pi z^2 c} \approx \frac{L_{edd}}{4 \pi z^2 c} \approx \frac{R_g c^2 m_p}{z^2 \sigma_T}
,\end{equation}
with $\displaystyle L_{edd} = 4 \pi G M m_p c /\sigma_T$ being the Eddington luminosity and $R_g = GM / c^2$ the gravitational radius.

This allows us to compare the inverse-Compton scattering time with the dynamical time of the system $t_{dyn} = z/c$ as
\begin{equation}
\label{eq:tic_tdyn}
\frac{t_{IC}}{t_{dyn}} = \frac{m_e}{m_p} \frac{z}{R_g} \frac{1}{\gamma}
.\end{equation}

\corrtwo{The bulk Lorentz factor of the plasma is the result of a balance of the radiative and dynamical forces. However, as shown here, in the inner parts of an AGN, the inverse-Compton scattering time is several orders lower than the dynamical time. In this case, a purely leptonic plasma is strongly tide to the photon field and its dynamic must be imposed by the inverse-Compton scattering as others forces (such as MHD forces or the interaction with the external flow) acting on timescales of the order of the dynamical time are too slow to counteract this force.}

% In the inner parts of an AGN, the inverse Compton scattering time is several orders lower than the dynamical time. In this case, a purely leptonic plasma is strongly tight to the photon field and its dynamic is then imposed by the external Compton radiation through the Compton rocket effect (see \citealt{1981ApJ...243L.147O}) as MHD forces acting in time-scales of the order of the dynamical time are too slow to counteract this force.

Therefore, the bulk Lorentz factor of the jet $\Gamma_{b}(z)$ is not a free parameter in the model but is imposed by the Compton rocket force and thus by the external photon fields.

 As shown in \cite{Vuillaume:2015jv} (hereafter Vu15), as long as most of the emission happens in the Thomson regime, the computation of the resulting $\Gamma_{b}(z)$ depends exclusively on the distribution of the external photon field in the inner parts of the jet whereas its final value $\Gamma_{\infty}$ reached at parsec scales \corr{when the bulk acceleration stops or becomes ineffective} depends on the jet energetics. Indeed, acceleration through the Compton rocket effect ceases when the scattering time in the rest frame of the plasma becomes greater than the dynamical time.
 
We use the method described in Vu15 to determine the equilibrium bulk Lorentz factor $\Gamma_{eq}(z)$ imposed by the Compton rocket and and only take into account the geometry of the external photon fields in the Thomson regime. We make the assumption that most of the inverse-Compton scattering always happens in the Thomson regime and we can verify this assumption afterwards for each object modelling.

% the evolution of $\Gamma_b(z)$ along the jet thanks to the external sources geometry described previously and to the physical conditions in the jet as long as equation \ref{eq:tic_tdyn} is verified. The complete calculation of...

From there, the evolution of $\Gamma_b(z)$ is determined by solving the differential equation (Vu15)
\begin{equation}
\frac{\partial \Gamma_b(z,\bar{\gamma})}{\partial z} = - \frac{1}{l(z,\bar{\gamma})} \left(\Gamma_b(z,\bar{\gamma}) - \Gamma_{eq}(z)\right)
,\end{equation}
with $l(z, \bar{\gamma})$ being the relaxation length to equilibrium that can be written as
\begin{equation}
l(z,\bar{\gamma}) = \frac{3 m_e c^3}{8 \pi \sigma_T} \frac{\beta_{eq}^3\Gamma_{eq}^3}{\bar{\gamma} H}\left(1+\frac{1}{3\Gamma_{eq}^2}\right)
,\end{equation}
with $\displaystyle H = \frac{1}{4\pi}\int I_{\nu_s} (\Omega_s)  \mu_s \dd \Omega_s \dd \nu_s$ being the Eddington parameter proportional to the net energy flux of the external photon fields on the jet axis.

In order to ease numerical computation, we do not solve the complete differential equation. The relaxation length is determined at each step and as long as $l(z,\bar{\gamma}) \ll z$ (equivalent to $t_{ic} \ll t_{dyn}$), one can consider that $\displaystyle \Gamma_b(z) = \Gamma_{eq}(z)$. At high $z$ in the jet, or when $\bar{\gamma}$ decreases enough, one has $l(z,\bar{\gamma}) \gg z$. In this case, jet moves in a ballistic motion and its speed reaches a constant $\Gamma_\infty$. Numerical tests showed that $\Gamma_{eq}(z)$ reaches $\Gamma_\infty$ when $l(z)/z \approx 0.6$. At this point we fix $\Gamma_b(z) = \Gamma_\infty$.

\corrtwo{The bulk Lorentz factor then reaches an asymptotic value corresponding to ballistic motion of the plasma. In the present model, we assume that the jet follows such a ballistic motion and is not affected by external factors (such as interstellar medium or two-flow interactions).}

\subsection{Absorption outside the jet \label{sec:abs_out}}

Between the jet and the observer, a photon encounters several radiation fields causing absorption through photon-photon interaction. There are two main sources of external radiation that should be considered here: the immediate vicinity of the jet and the extragalactic background light (EBL) (the absorption from the host galaxy being negligible).

\subsubsection{Extragalactic background light}
Extragalactic background light (EBL) is composed of several sources: the CMB, the diffuse UV-optical background produced by the integrated emission from luminous AGNs and stellar nucleosynthesis, and the diffuse infrared background composed of the emission from stars' photospheres and thermal emission from dust. Here we use absorption tables from \citep{Franceschini:2008em}. This introduces another term, $\exp(\tau_{ebl}),$ to the escape probability that depends on the source redshift $z_s$.

\subsubsection{Absorption in the jet vicinity}

The immediate vicinity of the jet is composed of several sources of soft photons.

\begin{figure}
\centering
\includegraphics[width=0.55\hsize]{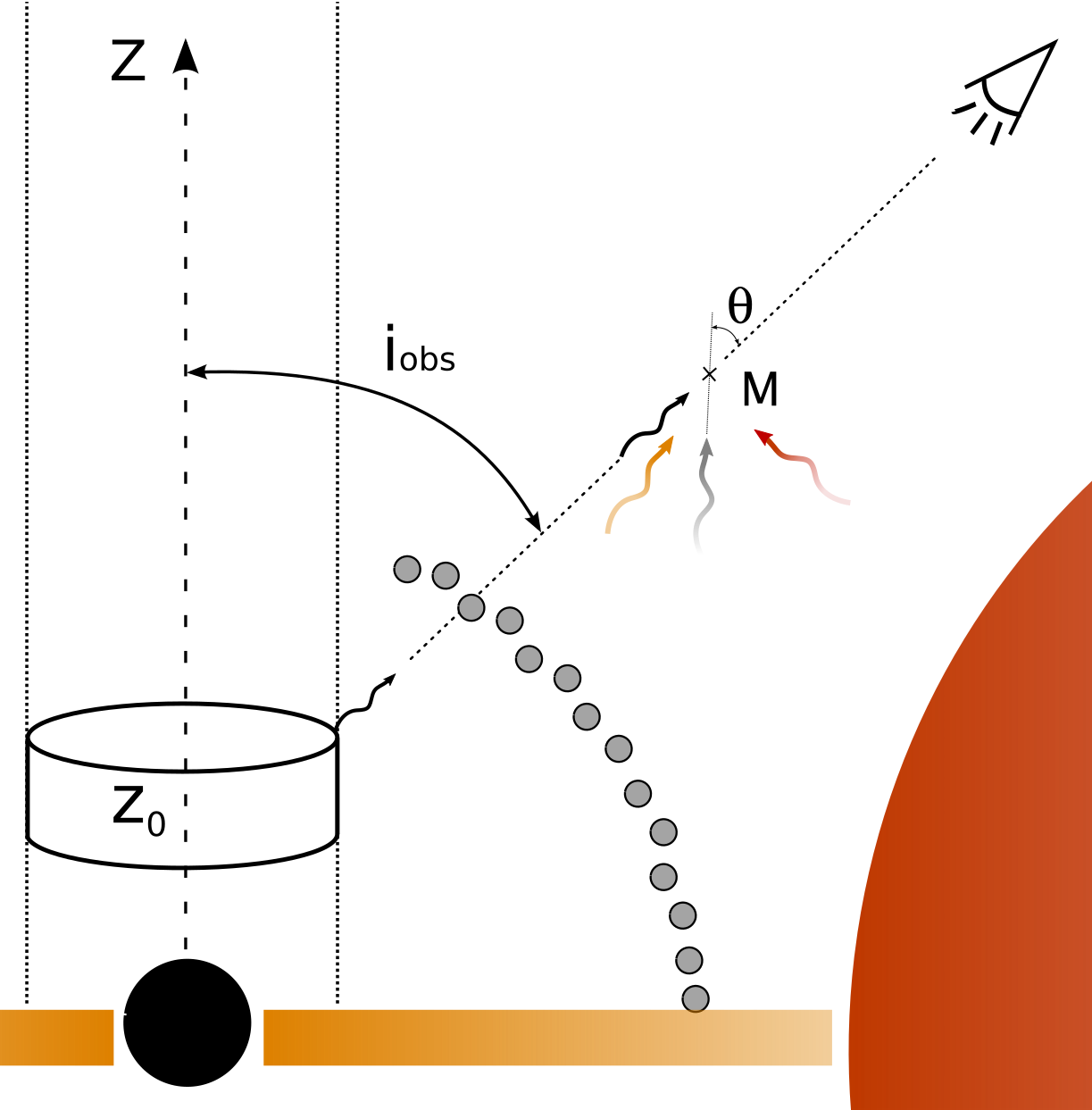}
\caption{\label{fig:absorption_sketch}
Sketch of the computation of the absorption from external sources. Photons from all external sources (accretion disc, BLR and torus) interact with photons emitted by the jet. For every emitting zone at $z_0$, one needs to integrate the absorption over the path of the gamma photons to the observer and over incoming directions of thermal photons.}
\end{figure}

\begin{figure}[h]
\centering
  \includegraphics[width=0.96\hsize]{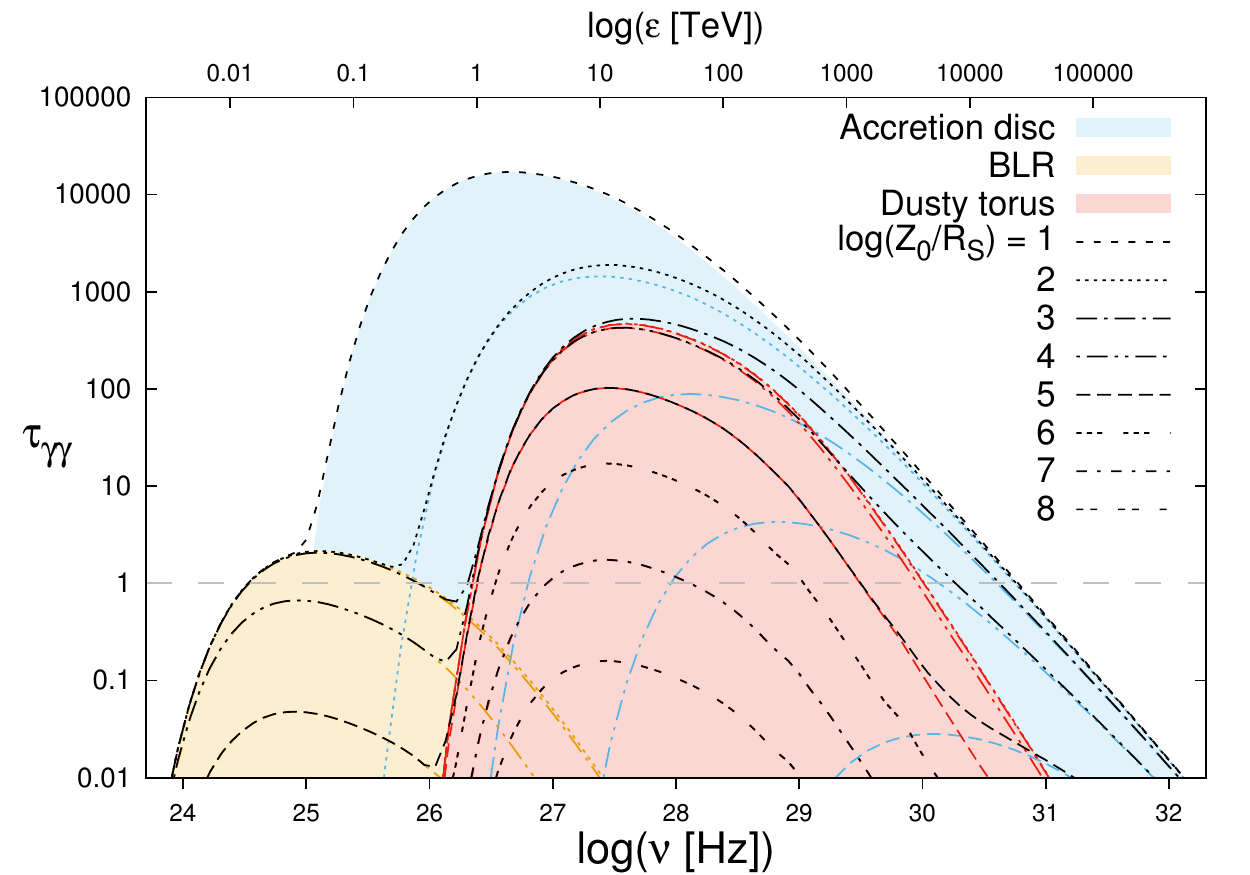}
  \caption{Integrated $\tau_{ext}$ created by the three external sources (disc, torus, BLR) and experienced by a photon of frequency $\nu$ leaving the jet from an altitude $z_0$ and travelling along the jet axis to infinity. Parameters of the sources:  \textit{Disc:} $L_d = 0.2 L_{edd}$, $R_{in} = 3R_S$ and $R_{out} = 5e4 R_S$. \textit{Torus:} $R_t = 5e4R_S$ and $D_t=1e5R_S$. \textit{BLR:} $R_{blr} = 8e3R_S$, $\omega \in [0:\pi/2]$, $\alpha_{blr}=0.01.$}
\label{fig:tau_sources}
\end{figure}

First, we consider the synchrotron photons from the jet itself. \cite{1995MNRAS.277..681M} showed that these photons interact on a typical distance corresponding to the jet radius $R$, which introduces an absorption term $\exp(-\tau_{jet})$ with $\tau_{jet}$ given by equation \ref{eq:tau_jet}. 

% Once obtained the opacity, one can calculate the escape probability $\displaystyle \Pp_{esc}^{jet}$  of photons produced in the jet. \citep{1995MNRAS.277..681M} showed that:

% \begin{equation}
%       \label{eq:escape_jet_marcowith}
%       \Pp_{esc}^{jet} = \left( \frac{1-\exp \left(-\tau_{jet} \right)}{\tau_{jet}}         \right) \exp \left( -\tau_{jet} \right)
% \end{equation}

% The left term in parenthesis is the usual solution of the transfer equation for photons absorbed in-situ. The extra exponential term has been introduced to account for the interaction of high-energy photons outside the jet with synchrotron radiation which is not confined to the jet on a characteristic distance $R(Z)$.

Once a photon has escaped the jet photon field, it enters the external photon field generated by external sources described in Sect. \ref{sec:ext_sources}. The induced opacity is generally greater than one in FSRQ and therefore cannot be neglected.
% Outside the jet, the pair production does not matter to our modelling but absorption from external sources can still be very important to the observed emission as the opacity can easily be greater than one in FSRQ.

As the external photon field resulting from the sources is highly anisotropic, one cannot use the approximation used in Sect. \ref{sec:abs_jet}. \corr{To compute the opacity, $\tau_{ext}$ , experienced by a photon emitted from the jet at an altitude $z_0$, one needs to integrate the complete $\gamma\gamma$ absorption given by equation \ref{eq:dTaugg} over the path followed by the photons from $z_0$ to infinity (here $M$ relates to the position along that path):
\begin{equation}
%\tau(z_0, \eps_1) = \int \d l(M) \int \frac{\d}{\d l}\tau_{\gamma\gamma}(\eps_1, M) \d \theta
\label{eq:abs_ext_particle_path}
\tau_{ext}(z_0, \eps_1) = \int \d l(M) \int \d \Omega(M) \int \d \eps_2 \frac{1-\cos\theta}{2} \, \sigma(\beta) \, n_{ph}(\eps_2, \Omega)
.\end{equation}

Terms in equation \ref{eq:abs_ext_particle_path} have been detailed at equation \ref{eq:dTaugg}. The factor $\theta(M)$ is the angle between the direction of the photon from the jet and the photons from thermal sources as illustrated in Fig. \ref{fig:absorption_sketch}. 
}

As an example, the result of this \corr{numerical integration, carried out for a range of energies and photon-emitted altitudes, is given in Fig. \ref{fig:tau_sources}. The figure represents isolines of photon-emitted altitude $z_0$.
In order to validate our approach, we chose an observing angle $i_{obs}=0$ and the same source parameters as in \cite{2007ApJ...665.1023R} and find consistent results.}\\
% for a photon travelling along the jet axis and are consistent with the results from \cite{2007ApJ...665.1023R}.

\corr{Considering the different sources of absorption described above,} the total escape probability of photons produced in the jet is then given by
\begin{equation}
        \label{eq:escape_tot}
        \Pp_{esc}^{tot} = \left( \frac{1-\exp \left(-\tau_{jet} \right)}{\tau_{jet}}         \right) \exp \left( -(\tau_{jet}+\tau_{ext}+\tau_{ebl}) \right)
,\end{equation}

and the specific intensity reaching the observer is thus given by $\displaystyle I_{obs}(\eps) = I_{jet}(\eps) \Pp_{esc}^{tot}(\eps, z_s).$

% \subsubsection{Influence of the observational angle on $\tau_{ext}$}

% \begin{figure}[h!]
% \begin{center}
%       \includegraphics[width=1\hsize]{figures/tau_sources_disc_08.pdf}
% \caption{Evolution of the absorption for several emission altitudes and angles}
% \label{fig:Taugg_iobs}
% \end{center}
% \end{figure}

% The photon field - and thus the absorption - perceived by an emitted photon depends not only on the emission site in the jet, but also on the emitting angle. One can therefore study the absorption coefficient as a function of the observation angle. In figure \ref{fig:Taugg_iobs}, we witness a strong difference between on-axis and off-axis observations. Off-axis, the absorption is both stronger and happens at lower energies. There are two effects at play here.

% \begin{itemize}
% \item The length of the photon path changes with the direction and is longer when the observational angle is greater.
% \item The angle between the jet radiation and the soft radiation changes $(\mu = \cos \theta)$ and is greater when the observational angle is greater. As the most efficient absorption happens for $\eps_1 \eps_2 = 4/(1 - \mu) $, when $\mu$ decreases (for a fixed soft radiation at $\eps_2$), higher energies photon $(\eps_1)$ are absorbed.
% \end{itemize}

\section{Example of application: the quasar 3C 273 \label{sec:3c273}}

3C 273 is the first quasar to have been identified thanks to its relatively close distance (redshift $z = 0.158$, \citealt{Schmidt:1963kd}) and has been extensively observed and studied. The broadband SED data used here come from \cite{1999A&AS..134...89T}; they averaged over 30 years of observations and are more likely to represent the average state of the AGN. This quasar represents a good test for the model as this average state can be associated to a quiescent state that we can model. Moreover, it is a FSRQ with relatively well-known external sources, allowing us to test the complete model with good constraints on these sources while imposing values of $\Gamma_b(z)$ through the Compton rocket effect (see section \ref{sec:gamma_b}). We detail our modelling of this source in the following section.

\subsection{Modelling}

% The accretion disc is well visible in the SED (figure \ref{fig:3C273_SED}) as the "big blue bump" in the optical band. For our modelling, we make a best-fit of this bump with the emission from a standard accretion disc \corr{from a Schwarzschild black-hole (see section \ref{sec:accrection_disc})}, thus imposing the black-hole mass at $ 1.3 \times 10^9$  solar mass and the accretion rate at $0.1$ Eddington accretion rate. \corr{The actual determining value being the absolute inner and outer radius of the accretion disc (see equation \ref{eq:disc_luminosity}), the same fit could be obtained with a more massive Kerr black-hole, the innermost stable circular orbit being relatively smaller.}\\

\corr{ The accretion disc is easily visible in the SED in Fig. \ref{fig:3C273_SED} as the big `blue bump' in the optical band. For our modelling, me make a best-fit of this bump (between $2.4 \times 10^{14}$Hz and $2.4 \times 10^{15}$Hz) considering the temperature given by a standard accretion disc (see equation \ref{eq:T_disc}) and with the two free parameters: $R_S = \frac{1}{3} R_{isco}$ and  $L_{disc}$ , the total luminosity of the disc. We obtain $R_S = 5.1 \times 10^{14}cm$ and $L_{disc} = 1.7 \times 10^{46}erg.s^{-1}$ ($\bar{\chi}^2 = 0.2$).
%$m = \dot{M} / \dot{M}_{edd}$ with $\dot{M}_{edd} = 48 \pi G M m_p / c\sigma_{Th} $ the Eddington accretion rate. We obtain $\dot{m} = 0.1$ and $R_{isco} = 2.3 \times 10^{15}cm$.
The modelling of the disc is given in orange in Fig. \ref{fig:3C273_SED}.}

The signature of the dusty torus is perceptible as a bump in the infrared band. Parameters of the torus are chosen to be in agreement with the position and luminosity of this bump. The resulting size of the torus is $R_T = 10^4 R_{S}$ and $D_T = 1.5 \times 10^4 R_{S}$. %\corr{A proper fit of the data would not make sense here as they result from the combination of the jet radio emission and of the torus emission.}

Strong emission lines are observed, indicating  the presence of a BLR. To model the BLR, we chose its radius from the value of \cite{Paltani:2005gk} that derived a size of $R_{blr}/c = 986$ days by studying the lag of the Balmer lines. The BLR opening is set to an ad hoc value $\omega_{max} = 35 \degree$ \corr{(there is no strong constraints on $\omega_{max}$, however, it must be large enough for the BLR to be outside the MHD jet)}. Its luminosity was set with $\alpha_{blr} = 0.1$ derived from observations \citep{Celotti:1997fi}.
% By studying the lag of the Balmer lines, \citep{Paltani:2005gk} find a BLR size of $R_{blr}/c = 986$ days. The size used in our modelling is smaller though with $R_{blr}/c = 75$ days.

\corr{
Emission attributed to a hot corona is observed in the SED of 3C 273 in the X-ray band. 
Following \cite{Haardt:1998tj} who used a model of thermal comptonization to reproduce the hot corona emission, we add an emission following a power law of photon index $\Gamma =1.65$ between $5\times10^{15}$ Hz and $5\times10^{19}$ Hz (0.02-200keV) as an extension of the accretion disc represented in Fig. \ref{fig:3C273_SED}. These photons contribute as well to the inverse-Compton emission and Compton rocket process.}

\corr{
Superluminal motion has been observed in 3C 273 by \cite{Pearson:1981gh}, proving the existence of relativistic motion in this jet. The deduced apparent velocity is $v_{app} \approx 7.5 c$\footnote{The authors compute an apparent velocity of $\sim 9.5c$ assuming a value for the Hubble constant $H_0 = 55 km.s^{-1}.Mpc^{-1}$ which is now known to be closer to $70 km.s^{-1}.Mpc^{-1}$}. This imposes constraints on the observation angle and one can determine that $i_{obs} < 15\degree$ (or $\cos i_{obs}$ > 0.96). Here we fixed $i_{obs} = 13\degree$ in agreement with these constraints.}

\corr{A best-fit of the model on the average data from \cite{1999A&AS..134...89T} is performed. Such a fit is a difficult task in this case as the model is computationally expensive and requires many parameters. We developed a method using a combination of genetic algorithms and gradient methods. As the source parameters have been fixed by observations, the free parameters of the model used for the fit are $R_0$, $n_0$, $B_0$, $Q_0$, $\lambda$, $\omega$ and $\zeta$.
 
% Because data points between $10^{16}$Hz and $2\times 10^{17}$Hz are thought to be due to the hot corona emission and not from the jet's, these points are not considered in the fit. The resulting reduced $\chi^2$ of the fit is then $\bar{\chi}^2 = 4.85$.

% All the model parameters are given in the table \ref{tab:3C273_param}. The resulting SED and parameters evolution are given in the figures \ref{fig:3C273_SED} and \ref{fig:3C273_params}. 
}

% Other parameters of the model given in the table \ref{tab:3C273_param} are the results of the SED fit on average data from \cite{1999A&AS..134...89T}. The resulting SED and parameters evolution are given in the figures \ref{fig:3C273_SED} and \ref{fig:3C273_params}. 

% The model reproduces well the data from radio to $\gamma$-rays. The different emitting processes or sources are represented with different colors while the jet is sliced in several zones whose emission is represented with different dotted lines.

% \begin{table}[ht]
% \centering
% \begin{tabular}{c c | c c | c c  }
% \toprule
% $M_\bullet/ M_\odot$& $1.3\times10^9$ & $\dot{m}$ & 0.1 & z & 0.158  \\
% $R_S$(cm) & $3.8\times10^{14}$& $\omega_{max} $ & 35\degree & $\alpha_{blr} $ & 0.1\\
% $R_{blr}/R_S$  & $6.5\times 10^3$ & $R_T/R_S$ & $10^4$ &  $ R_{in}/R_S $ & 3\\
% $D_T/R_S$ & $1.5 \times 10^4$  & $i_{obs}$ & 13\degree & $R_0/R_S$ &7.5  \\
% $R_{out}/R_S$ & $5 \times 10^3$ & $B_0$ & 12 G & $Q_0$ & $0.03 s^{-1}$\\
% $Z_0/R_S$ & $2\times 10^3$ & $Z_c/R_S$& 1$0^9$ & $\omega$ & 0.5  \\
% $n_0(cm^{-3})$ & $4.5 \times 10^3$ & $\lambda$ & 1.4 & $\zeta$ & 1.52 \\
% \bottomrule
% \end{tabular}
% \caption{\label{tab:3C273_param}Parameters corresponding to 3C 273 modelling.}
% %\corr{table to rearrange so it look better}}
% \end{table}

\begin{table}[ht]
\centering
\resizebox{\columnwidth}{!}{%
\begin{tabular}{c c | c c | c c  }
\toprule
$R_{S}$(cm) & $5.3\times10^{14}$& $Z_c/R_S$& 1$0^9$ & $L_{disc} (erg/s)$ & $ 1.7 \times 10^{46}$\\
$ R_{in}/R_{S} $ & 3 & $R_0/R_S$ & $7.5$ & $n_0 (cm^{-3})$ & $4.5 \times 10^3$ \\
$R_{out}/R_{S}$ & $5 \times 10^3$ & $\omega_{max} $ & $35\degree$ & $B_0$ & 12 G\\
$D_T/R_{S}$ & $1.5 \times 10^4$  &  $\alpha_{blr} $ & 0.1 & $Q_0$ & $0.03 s^{-1}$ \\
$R_T/R_{S}$ & $10^4$ & $i_{obs}$ & 13\degree & $\lambda$ & 1.4 \\
$R_{blr}/R_{S}$ & $4.8 \times 10^3$ &  z & 0.158 & $\omega$ & 0.5  \\
$Z_0/R_{S}$ & $2 \times 10^3$ & &  & $\zeta$ & 1.52 \\
\bottomrule
\end{tabular}%
}
\caption{\label{tab:3C273_param}Parameters corresponding to 3C 273 modelling.
$R_S = 2GM/c^2$ is the Schwarzschild radius deduced from a best-fit of the optical emission. $R_{in}$ and $R_{out}$ are the inner and outer radii of the disc. $D_T$ and $R_T$ are geometrical parameters of the torus. $R_{blr}$ is the BLR radius. $Z_0$, $R_0$ , and $Z_c$ are geometrical parameters of the jet (see section \ref{sec:jet_geometry}). $\omega_{max}$ corresponds to the BLR opening and $\alpha_{blr}$ to its luminosity fraction with respect to that of the disc, $L_{disc}$. The value z is the redshift of 3C273. The factors $n_0$, $B_0$, $Q_0$, $\lambda$, $\omega$ and $\zeta$ are free parameters of the jet model as described in section \ref{sec:jet_geometry}}
\end{table}

\begin{figure}[ht]
\begin{center}
        \includegraphics[width=\hsize]{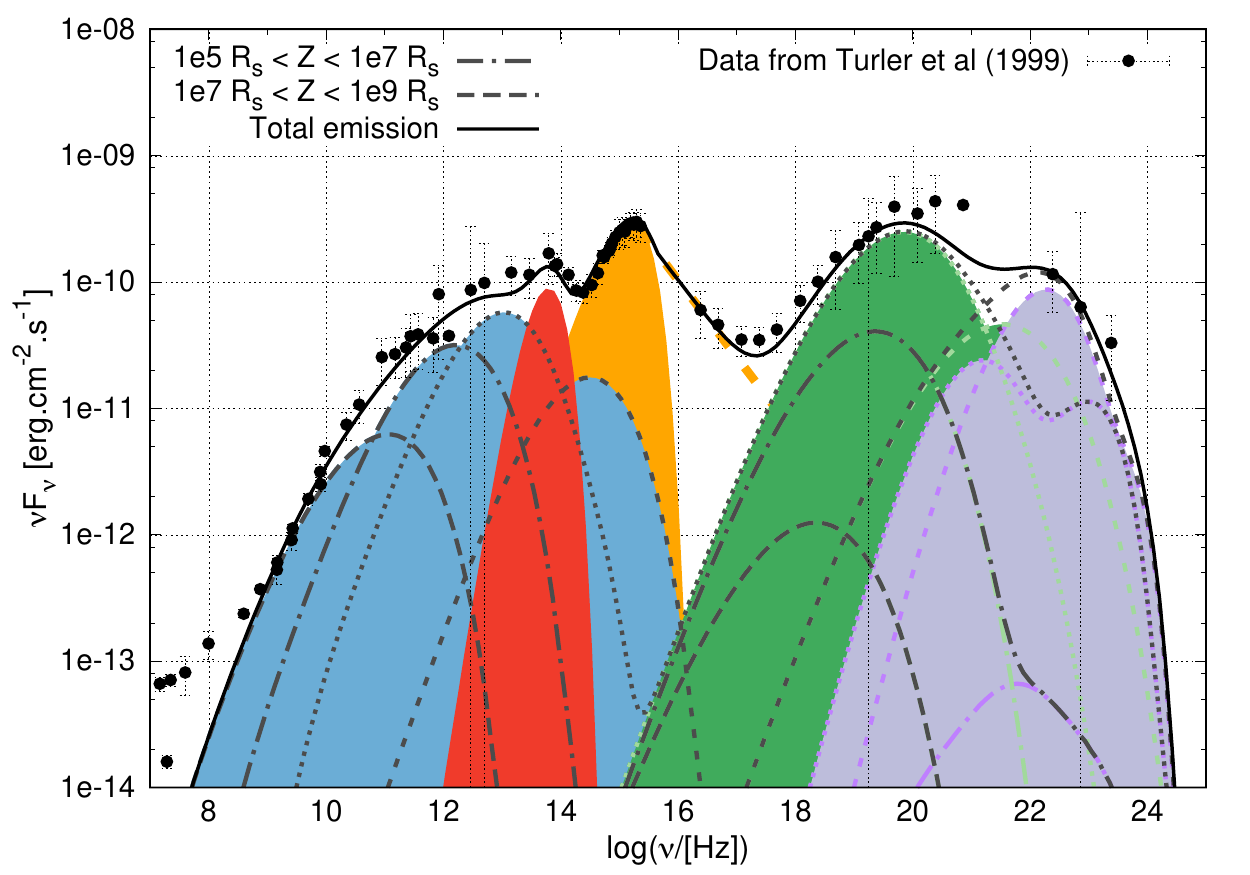}
\caption{SED of 3C 273 - modelling compared to data from \cite{1999A&AS..134...89T}. The synchrotron emission is shown in blue, the SSC in green, and the external Compton in purple. The torus emission is shown in red and the multicolor accretion disc in orange (filled orange curve), complete with a power-law describing the hot corona emission between 0.02 and 200 KeV (dashed orange line). Different emission zones in the jet are represented with different dotted lines. The emission below $10^9$Hz not reproduced by the model is interpreted as the emission from the jet hot spot, not modelled here.}
\label{fig:3C273_SED}
\end{center}
\end{figure}

\begin{figure}[ht]
\centering
  \includegraphics[width=1\hsize, trim = 0cm 0cm 0cm 1cm, clip=true]{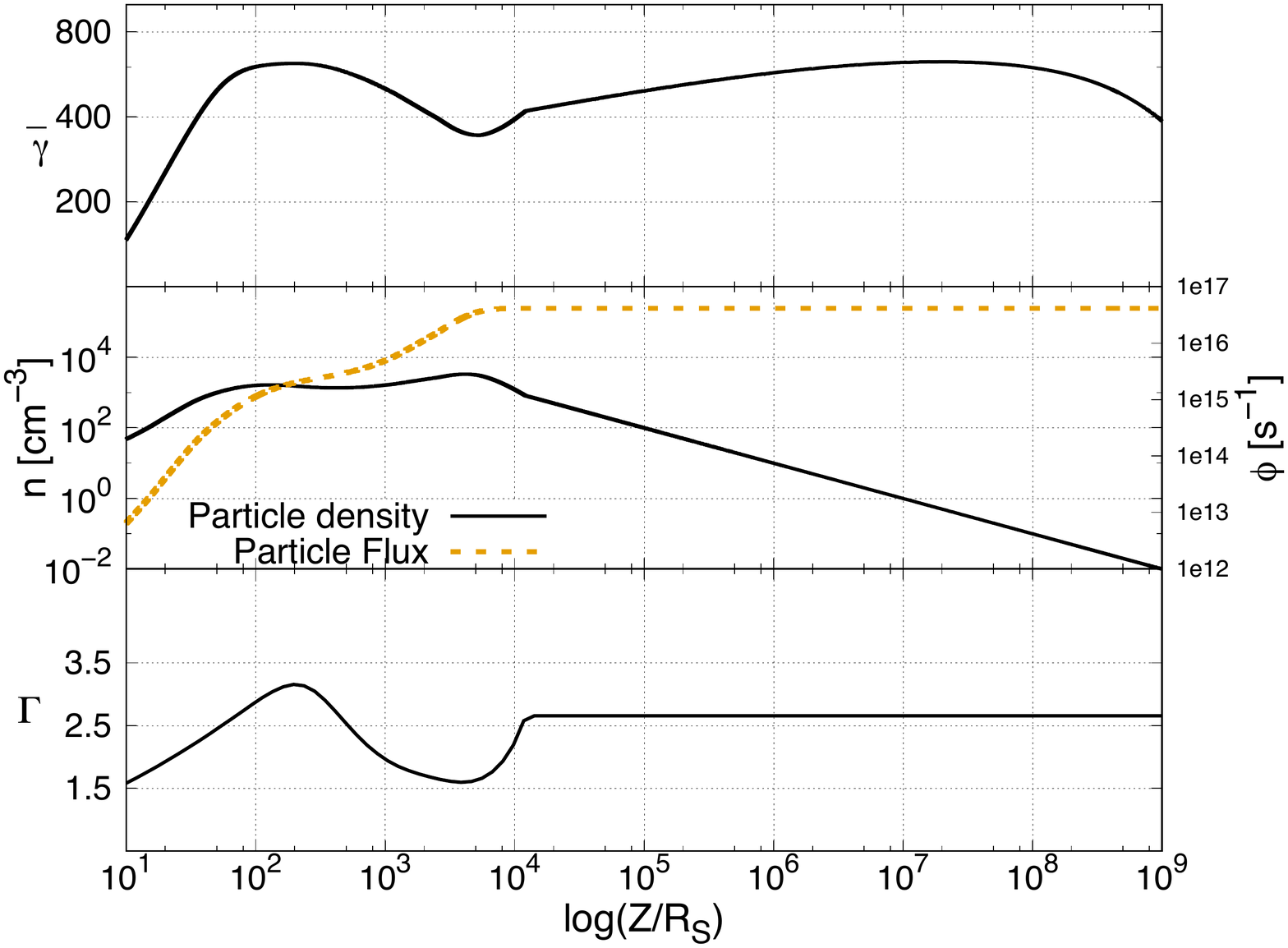}
  \includegraphics[width=1\hsize, trim = 0cm 0cm 0cm 1cm, clip=true]{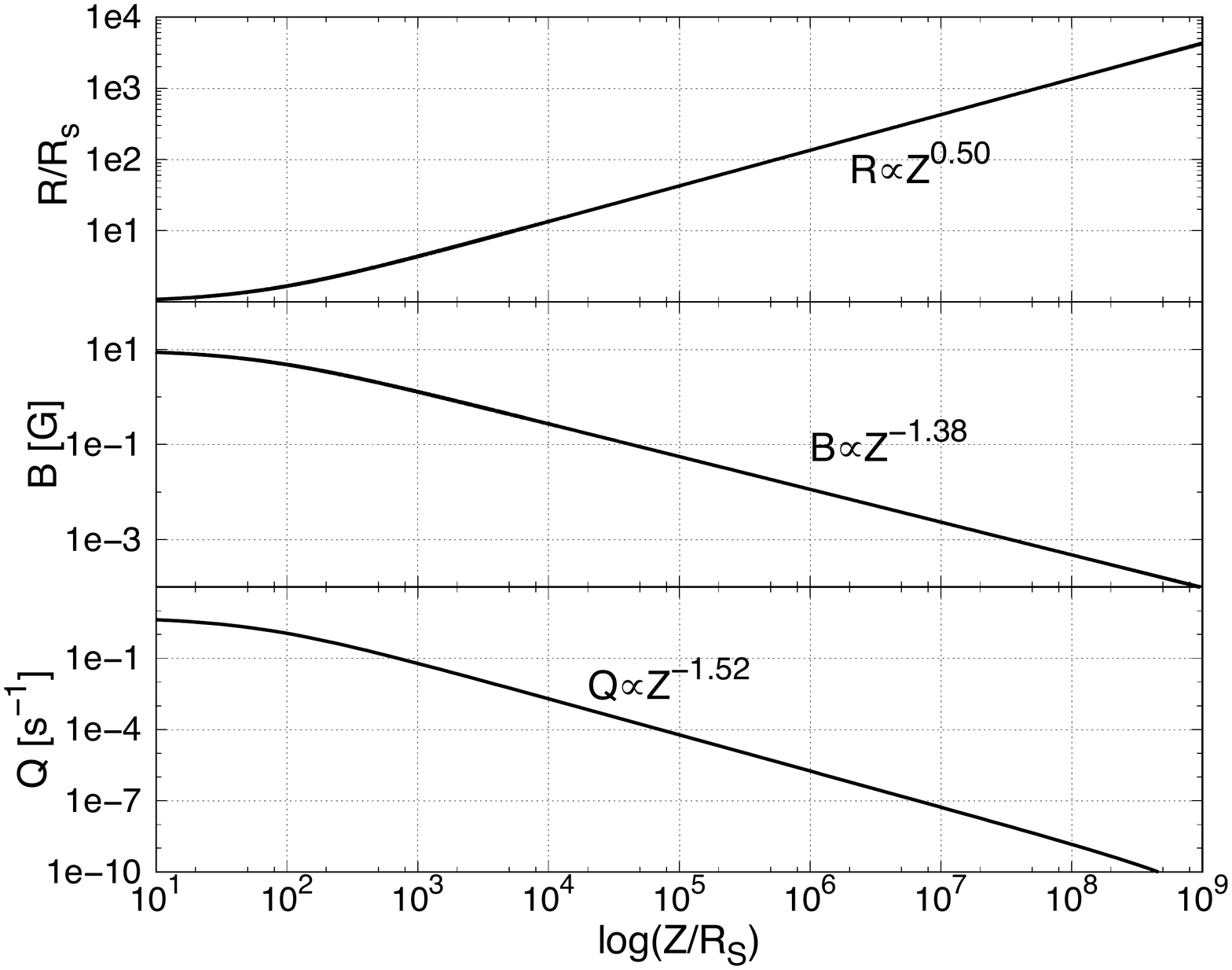}
  \includegraphics[width=1\hsize, trim = 0cm 0cm 0cm 1cm, clip=true]{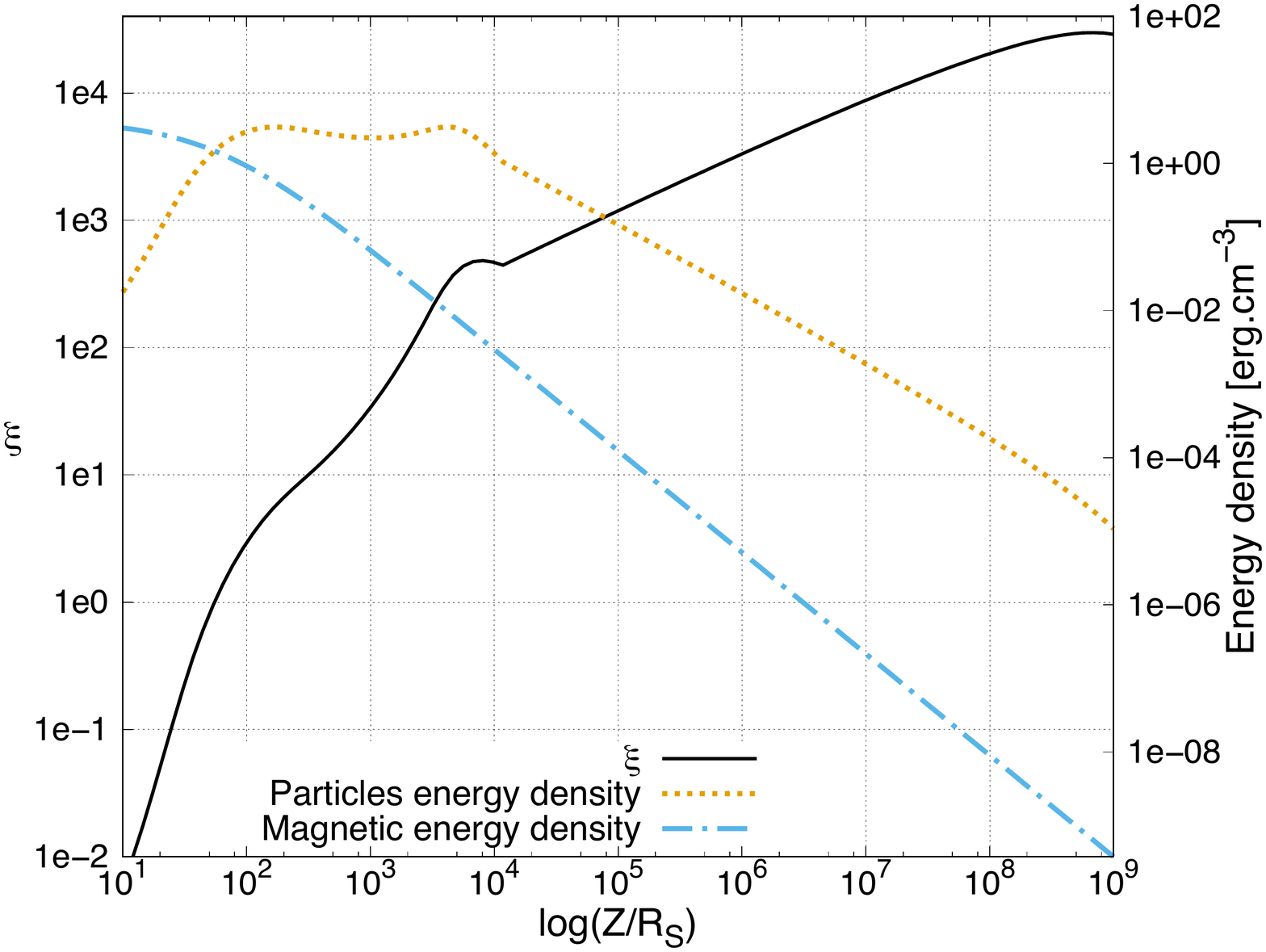}
\caption{Evolution of the physical parameters as a function of the distance in the jet for the modelling of 3C273}
\label{fig:3C273_params}
\end{figure}

\subsection{Results and discussion}

% Fitting the optical spectrum with the emission from a standard accretion disc around a Schwarzschild black-hole can help fixing the black-hole mass and the accretion rate.

\corr{ With the seven free parameters, the fit gives a reduced $\bar{\chi}^2 = 4.92$. The model accurately reproduces the entire SED of 3C273 from radio to gamma-rays except for the far radio below $10^8$Hz. These points show a break in the spectrum and are interpreted as the jet hot spot on a very large scale. This hot spot is not supposed to be modelled here and it is actually not surprising that these points are not well fitted by the best fit model; if we do not take them into account, then the reduced $\chi^2$ is reduced to  $\bar{\chi}^2 = 1.3$, more representative of the quality of the fit obtained.}\\

\corr{
The mass and accretion rate of the black-hole can be deduced from the radius $R_{S}$ and the luminosity $L_{disc}$ determined previously, depending on the BH spin. For a Schwarzschild (non-spinning) black-hole, $M_\bullet = R_{isco} c^2/ 6G = 1.8 \times 10^{9} M_\odot$ and the reduced accretion rate $\dot{m} = \dot{M}/\dot{M}_{edd} = 0.08$.
For a maximum spinning Kerr black-hole however, $M_\bullet = R_{isco} c^2/ 2G = 5.4 \times 10^9 M_\odot$ and $\dot{m} = 0.027$. These values of mass are in agreement with those derived from observations;  \cite{Paltani:2005gk} determined $M_\bullet = (5.69 - 8.27) \times 10^9 M_\odot$ using the reverberation method on Ly-$\alpha$ and $C_{IV}$ lines and $M_\bullet = (1.58-3.45) \times 10^9 M_\odot$ using Balmer lines whereas \cite{Peterson:2004ig} determined a mass of $M_\bullet = (8.8 \pm 1.8) \times 10^9 M_\odot$ using a large reverberation-mapping database.}\\

% Concerning the standard accretion disc and the black-hole inferred mass of $M_\bullet =  1.3 \times 10^9 M_\odot$, one can notice that it is lower than the ones derived by observation - \cite{Paltani:2005gk} determined $M_\bullet = (5.69 - 8.27) \times 10^9 M_\odot$ using reverberation method on Ly-$\alpha$ and $C_{IV}$ lines and $M_\bullet = (1.58-3.45) \times 10^9 M_\odot$ using Balmer lines whereas \cite{Peterson:2004ig} determined a mass of $M_\bullet = (8.8 \pm 1.8) \times 10^9 M_\odot$ using large reverberation mapping database. The lower mass found here could be explained by the fact that we used a non rotating Schwarzschild black hole differing from a rotating Kerr black hole that would imply a larger mass for the same inner radius. However, only the photon flux and angular distribution is important here and this distinction is largely immaterial for our model as long as the observed flux is fitted.  

In the jet, the low energy (radio to optical) is produced by the synchrotron process. The synchrotron part of the inner jet (below $10^3 R_S$) emits in the optical and is hidden by the emission from the accretion disc and the dusty torus. When moving further in the jet, the synchrotron peak shifts to lower frequencies and the further we go, the more the peak shifts. Finally, the whole jet from $10^3 R_S$ to $10^9 R_S$ is necessary to reproduce the power-law-like radio spectrum. Its slope is determined by different factors: the increase of the jet radius, the decrease of the magnetic field and of the particle heating, and the bulk Lorentz factor. The spectrum of 3C 273 shows a break at $\sim 10^9$ Hz which is poorly fitted by our model and is thought to be the result of synchrotron emission produced by the extended radio structure (lobe+hot spot). \corrtwo{It could also show the limitation of our assumption on ballistic motion of the jet at very large distances. A deceleration of the jet could result in the observed break in the spectrum.}\\

The high-energy emission is produced by inverse-Compton processes, either on synchrotron (SSC) or on thermal photons (EC). Similarly to the synchrotron, the highest energies are produced close to the central engine and further regions emit at lower energies. In particular, the spectrum at $\nu > 10^{21}$ Hz is produced by regions at $z < 10^3 R_S$ with a combination of SSC and EC. However, the X-rays and soft $\gamma$-rays are produced further, at $z > 10^3 R_S$ by SSC.\\

Figure 8 shows the evolution along the jet of the different model parameters. One can see first that the jet reaches ballistic motion ($\Gamma_b(z) = \Gamma_{\infty}$) at $z = 10^4 R_S$. From this point, there is no further pair creation, the density decreasing only by dilution (due to the increase of the jet radius), and all parameters follow a very smooth evolution, corresponding to the almost featureless spectrum below $10^{13}$Hz.\\

Concerning the geometrical parameters, the spine geometry (see equation \ref{eq:Rz}) follows a parabola with a radius $R(z) \propto \sqrt{z}$. The jet opening can be evaluated by $\displaystyle \tan \theta_{jet} = \frac{R}{z}$  and goes from $2\times10^{-3}$rad below 1 parsec to about $10^{-5}$rad at 1kpc which makes it a very narrow jet (these values concern only the spine jet here which is collimated by the outer MHD jet in the two-flow paradigm). However, the jet radius is important only to compute the SSC radiation, the other process depending only linearly on the number of particles: if the jet were to widen after the SSC emitting region, this would not affect the overall spectrum.\\

The particle density (middle curve of the top plot in the Fig. \ref{fig:3C273_params}) evolves along the jet and goes through important pair creation under $10^3 R_S$. This part is also the part where most of the high-energy emission is emitted. This is in good agreement with the two-flow paradigm as flares (which can be extreme at high energies) are explained by intense phases of pair creation. Slight changes in the particle acceleration could induce large changes in the pair creation which is a highly non-linear process, in turn creating flare episodes. A time-dependent model such as the one described in \cite{Boutelier:2008bga} is currently used to describe these flares.\\

The particles mean Lorentz factor ($3\bar{\gamma}$) varies between 600 and 1800; it increases and decreases rapidly below $10^3 R_S$, following changes in the bulk Lorentz factor $\Gamma_b$. This is due to changes in the cooling inferred by changes in Doppler aberration in the plasma rest frame and in the particle density along the jet. Further in the jet, $\bar{\gamma}$ slowly increases but due to an important decrease in the particle density, the particle energy density also decreases (dotted line in the last plot in Fig. \ref{fig:3C273_params}).\\

The last plot in Fig. \ref{fig:3C273_params} represents the energy density in the particles and in the magnetic field as well as the equipartition ratio defined as the ratio of both energy densities:
\begin{equation}
        \xi = \frac{n_e <\gamma>m_e c^2}{B^2/8\pi}
\label{eq:xi}
.\end{equation}

This plot shows that the jet is highly magnetised very close to the black-hole (below 100 $R_S$) as much of the energy is carried by the magnetic field. Particles then carry much more overall energy and bring this energy very far along the jet, allowing for the far synchrotron emission.\\

The bulk Lorentz factor $\Gamma_b$ varies as a function of the distance in the jet and is equal to $\Gamma_{eq}(z)$ under $10^4 R_S$. Changes in $\Gamma_b(z) = \Gamma_{eq}(z)$ are due to effects discussed in \cite{Vuillaume:2015jv} and are the result of the Compton rocket effect. The jet reaches ballistic motion at $z = 10^4 R_S$ \corr{as the acceleration becomes ineffective and then $\Gamma_b(z)$ reaches its final value $\Gamma_\infty$.}\\

The value of $\Gamma_\infty \approx 2.7$ is very questionable in this modelling, as superluminal motion inferring $v_{app} > 7c$ has been observed, implying at least the same value for $\Gamma_b$. Superluminal motion as seen by very long base interferometry (VLBI) corresponds to regions very far from the central engine and should thus be produced by components travelling at $\Gamma_\infty$. Therefore, $\Gamma_\infty$ should be able to reach higher values. In the Compton rocket model, this implies that particles should be coupled to the photon field at larger distances than what is found here (well outside the BLR region). There are two simple ways to explain this lack of energy in particles:%\LEt{please remove the following bullet system and number the two ways (1) and (2) if necessary.}
\begin{enumerate}
\item
our description of the acceleration with a power-law might be too simple. It is quite clear that jets are more complex objects and that some re-acceleration sites are present far in the jet, as shown by the complex images of the jet (see also for example HST-1 in M87). Such far acceleration would give an energy boost to particles that would turn into jet acceleration on greater scales, pushing $\Gamma_\infty$ to the observed values.
\item
Our modelling of 3C 273 corresponds to a quiescent state. During flaring states, it is expected that more energy is transferred to particles, here again providing a possibility to reach higher values of $\Gamma_\infty$. In this view, the fastest features observed in jets would correspond to past flaring episodes that accelerated the plasma to high values of $\Gamma_b$ - the latter now following ballistic motion.
\end{enumerate}

On the other hand, it is very interesting that the model is able to reproduce the broadband spectrum of 3C 273 with $\Gamma_b(Z) < 3$.
As explained in section \ref{sec:jet_velocity}, there is substantial evidence pointing to low Lorentz factors in AGNs and high Lorentz factors are very difficult to explain if applied to the entire jet. Here we show that with a stratified jet model, high Lorentz factors are not necessary to reproduce the averaged broadband emission and that the quiescent state of an object could correspond to very moderate $\Gamma_b$ in opposition to episodes of flares that could induce higher $\Gamma_\infty$, as high as the ones observed.

%\section{Discussion}

%\corr{
%The two-flow paradigm is a very coherent model able to explain most of AGN jets features. It must be considered as a whole paradigm as it differs from the generally accepted view considering localised sources of emission powered by shocks. Moreover, every piece of the model is essential and works in synergy with the rest...
%}

\section{Conclusion}

Emission modelling is essential to understanding the AGN jet physics, and yet despite many flaws the one-zone model is still predominant. However, more complex models are being developed, including structured jets models. These models introduce more physics and improve our understanding of AGNs as they are often better at explaining the observations. In this work, we present a numerical model of a stratified jet based on the two-flow idea first introduced by \cite{1989MNRAS.237..411S} and further developed by \cite{1991ApJ...383L...7H}. 

The numerical model is based on a prescription of the geometry of the jet providing the evolution of the radius, magnetic field, and particle acceleration. External thermal sources such as the accretion disc, the dusty torus, and the broad-line region are also modelled, including their spatial geometry.
In the jet, we consider emission from synchrotron and inverse-Compton (over synchrotron and thermal photons) processes. The physical conditions such as the particles energy and density are consistently computed along the jet based on initial conditions at a fixed point. In particular, the bulk Lorentz factor is not a free parameter in our model but its evolution is entirely constrained by the external sources (accretion disc, dusty torus, BLR) through the Compton rocket process.
Photon-photon absorption is taken into account in the jet (essential for pair production) as well as outside the jet, in its vicinity, and on the photon path to the observer. 

Despite being very constrained, the model is able to accurately reproduce the broadband emission of 3C273 from radio to gamma-rays. As the physical conditions evolve along the jet, different zones contribute to different parts of the observed spectrum. Inner parts of the jet contribute more to high-energies while low-frequency radio is emitted further in the jet, at parsec scales.

In the future, the model could be extended to incorporate variability (as done in \citealt{Boutelier:2008bga}) and can be applied to other types of compact object such as galactic X-ray binaries. Further work is being planned for these applications.

\section*{Acknowledgement}
We acknowledge funding support from the French CNES agency and CNRS PNHE. IPAG is part of Labex OSUG@2020 (ANR10 LABX56).

\bibliographystyle{aa}
\bibliography{biblio}

\begin{appendix}

\section{Computation of the synchrotron and synchrotron-self Compton radiation}

\subsection{Synchrotron radiation}

The function $\Lambda$ presented in equation \ref{eq:Jsyn} can be approximated analytically in the different regimes:
\begin{itemize}
\item
A development in Taylor series for small $y$ gives:
\begin{equation}
        \Lambda(y) = \frac{8 \pi}{9 \sqrt{3}} \left( \frac{y}{2} \right)^{-2/3} \qquad \text{for}\quad y \ll 1
    \label{eq:lambda1}
.\end{equation}

\item
\citep{Mahadevan:1996ce} proposed an approximation for intermediates values of $y$:
\begin{equation}
        \Lambda(y) = 2.5651 \left( 1 + \frac{1.8868}{y^{1/3}} + \frac{0.9977}{y^{2/3}} \right) \exp \left( -1.8899 y \right)
    \label{eq:lambda2}
.\end{equation}

\item
For large values of $y$, with a saddle-point method, \citep{Sauge:2004tc} obtains:
\begin{equation}
        \Lambda(y) = \frac{2 \pi}{\sqrt{6}} \exp \left [ -\left( 2^{1/3} + 2^{-2/3}\right)y^{1/3}\right ] \qquad \text{for}\quad y \gg 1
    \label{eq:lambda3}
.\end{equation}

\end{itemize}

\subsection{Synchrotron-self Compton radiation \label{sec:appendix_ssc}}

In this section, we consider the dimensionless energy (as introduced before) $\displaystyle \eps = h\nu/m_e c^2$. The photon production rate resulting from the scattering of a synchrotron photon field $n_{ph}(\eps)$ on a population of particles $n(\gamma)$ cnan be written as:
\begin{equation}
\label{eq:compton_rate}
\frac{\d n (\eps_1)}{\d \eps_1 \d t} = \iint \d \eps \d \gamma \frac{\d n_{ph}}{\d \eps} \frac{\d n}{\d \gamma} K_{jones}(\eps_1, \eps, \gamma)
,\end{equation}
with $K_{jones}$ being the Compton kernel for an isotropic source of soft photons, considering the full Klein-Nishina cross section in the head-on approximation as evaluated by \cite{1968PhRv..167.1159J}. Numerical evaluation of this equation is time consuming but approximations may be used to speed up calculations. To do so, we consider the Thomson and the Klein-Nishina regimes separately:
% \begin{equation}
% \label{eq:k_jones}
% K_{jones} = \frac{3}{4} \frac{c \sigma_{Th}}{\eps \gamma^2}f(q, \zeta_\eps) \Phi(q -1/4\gamma^2)\Phi(1-q)
% \end{equation}

\paragraph{\#} Emission in the Thomson regime\\
In this regime, \cite{BLUMENTHAL:1970gb} and \cite{1979rpa..book.....R} showed that the Jones kernel can be simplified:

\begin{equation}
\label{eq:k_jones_thomson}
K_{jones}\left( \eps_1, \eps, \gamma \right) = \frac{3 \sigma_{th}}{4 \eps \gamma^2} f \left( \frac{\eps_1}{4 \eps \gamma^2} \right)
,\end{equation}
with $\displaystyle  f(x) = 2 x\ln x + x + 1 - 2x^2$.\\

Injecting this simplified kernel into equation \ref{eq:compton_rate} and considering a pile-up distribution (see equation \ref{eq:pileup}) for the particles, one obtains
\begin{equation}
\frac{\d n^{Th}}{\d \eps_1 \d t} = \frac{3}{4} c \sigma_{Th} \int \frac{\d \eps}{\eps} \frac{\d n_{ph}}{\d \eps} F(\eps,\eps_1)
,\end{equation}
with
\begin{equation}
F(\eps,\eps_1) = n_0 \bar{\gamma} \int_{1/\bar{\gamma}}^{\infty} \d u e^{-u} f \left( \frac{1}{u^2} \frac{\eps_1}{4 \bar{\gamma}^2 \eps_0}\right)
.\end{equation}

Using the approximation for the function $f$ proposed by \cite{1979rpa..book.....R} $\displaystyle f(x) \approx 2(1-x)/3$, one can show that the integral can be analytically solved and gives
\begin{equation}
F(\eps, \eps_1) = n_0 \bar{\gamma} \tilde{g}\left( \frac{\eps_1}{4 \bar{\gamma}^2 \eps_0}\right)
,\end{equation}
with
\begin{equation}
\begin{aligned}
& \tilde{g}(x) = \frac{2}{3} e^{-\sqrt{x}} \left( 1 - \sqrt{x} + x e^{\sqrt{x}} \Gamma(0, \sqrt{x}) \right) \\
& \Gamma(a,x) = \int_x^\infty e^{-t} t^{a-1} \d t
\end{aligned}
,\end{equation}
where $ \Gamma(a,x)$ is the incomplete gamma function introduced by \cite{Abramowitz:1974:HMF:1098650}.

Finally, the SSC photon production rate in the Thomson regime can be written as
\begin{equation}
\label{eq:dn_th}
        \frac{\d n^{Th} (\eps_1) }{\d t} = \frac{3}{4} \sigma_{Th} \: \frac{N_e}{2 \bar{\gamma}^2} \: \tilde{G}\left( \frac{\eps_1}{\bar{\gamma}^2}\right)
,\end{equation}

 with the function 
\begin{equation}
\begin{aligned}
        \tilde{G}(x) = & \int \frac{\d n_{ph}}{\d \eps} \frac{2}{3} \exp\left(-\sqrt{\frac{x}{4\eps}} \right) \\
    & \times \left[ 1 - \sqrt{\frac{x}{4\eps}} + \frac{x}{4\eps} \exp\left(\sqrt{\frac{x}{4\eps}}\right) E_i \left( \sqrt{\frac{x}{4\eps}} \right)       \right] \frac{\d \eps}{\eps}
 \end{aligned}
 ,\end{equation}

and the exponential function
\begin{equation}
        E_i (x) = \Gamma(0,x) = \int_x^{\infty} \frac{\exp(-t)}{t} \, \d t
.\end{equation}

As one can see, in equation (\ref{eq:dn_th}), the function $\tilde{G}$ depends only on the ratio $\eps_1/\bar{\gamma}^2$. This function can therefore be evaluated and tabulated to decrease the computing time.

\paragraph{\# Emission in the Klein-Nishina regime}

In this regime, the complete Jones kernel must be considered.
However, \cite{Rieke:1969eo} showed that some approximations were possible, introducing the differential cross-section $\sigma(\eps_1, \eps_0, \gamma)$, one can write
\begin{equation}
K_{jones} = c \sigma(\eps_1, \eps_0, \gamma) = c \sigma_{Th} f(\eps_0, \gamma) \delta(\eps_1 - <\eps_1>)
,\end{equation}
 in the Thomson (1) and the Klein-Nishina (2) regimes, respectively,
\begin{equation}
\begin{aligned}
(1) \, & f(\eps_0, \gamma) = 1 \quad \text{and} \quad <\eps_1> = \gamma^2\eps_0 \\
(2) \, & f(\eps_0, \gamma) = \frac{3}{8} \frac{\ln(2\gamma\eps_0) +1/2}{\gamma \eps_0} \quad \text{and} \quad <\eps_1> = \gamma
\end{aligned}
.\end{equation}

Then the photon production rate can be written as
\begin{equation}
\label{eq:ssc_rate_appendix_gi}
\frac{\d n^{kn}}{\d \eps_1 \d t} = c \sigma_{Th} \int \d \eps_0 \frac{\d n_{ph}}{\d \eps_0} g_i(\eps_1, \eps_0)
,\end{equation}
where $g_i$ must be calculated in the two diffusion regimes.\\

In the Thomson regime, one can show easily that for a pile-up distribution,
\begin{equation}
g_{Th}(\eps_1, \eps_0) = \frac{n_0}{2} \eps_0^{-3/2} \eps_1^{1/2} \exp\left( -\sqrt{\frac{\eps_1}{\eps_0 \bar{\gamma}^2}} \right)
.\end{equation}
Injecting this expression into \ref{eq:ssc_rate_appendix_gi} and assuming a power law for the soft photon spectrum $\displaystyle 
\d n_{ph}/\d \eps = n_s \eps^{-s}$ over the range $[\eps_{min}, \eps_{max}]$  after some manipulations one obtains
\begin{equation}
\frac{\d n}{\d \eps_1 \d t} = n_s n_0 c \sigma_{Th} \eps_1^{-s} \bar{\gamma}^{2s+1} \left(\Gamma(2s+1, u_{min}) - \Gamma(2s+1, u_{max}) \right)
,\end{equation}
with $\Gamma$ the incomplete gamma function and
with $\displaystyle u_{min} = \sqrt{\frac{\eps_1}{\eps_{max}\bar{\gamma}^2}}$ and $\displaystyle u_{max} = \sqrt{\frac{\eps_1}{\eps_{min}\bar{\gamma}^2}}. $

In the Klein-Nishina regime, with a pile-up distribution injected into \ref{eq:ssc_rate_appendix_gi}, one obtains
\begin{equation}
g_{kn} (\eps_1, \eps_0) = \frac{3 n_0}{8} \eps_1 \frac{\ln(2\eps_1\eps_0) +1/2}{\eps_0} \exp\left( -\frac{\eps_1}{\bar{\gamma}}\right)
.\end{equation}
With the assumption of a power law for the soft photon spectrum, the production rate becomes
\begin{equation}
\begin{aligned}
\frac{\d n}{\d \eps_1 \d t} & = 
\frac{3}{8} n_s n_0 c \sigma_{Th} \eps_1 \exp\left(-\frac{\eps_1}{\bar{\gamma}}\right) \\
& \times \left\{ \left(\ln(2\eps_1) +\frac{1}{2} \right) K_1^{(s)}(1/\eps_1, \eps_0) +  K_2^{(s)}(1/\eps_1, \eps_0) \right\} 
\end{aligned}
,\end{equation}

with
\begin{equation}
\begin{aligned}
& K_1^{(s)}(\eps_{min}, \eps_{max}) = \left[ \frac{x^{-s}}{s}\right]^{\eps_{max}}_{\eps_{min}} \\
& K_2^{(s)}(\eps_{min}, \eps_{max}) = \left[ \frac{x \exp[-(s+1)\ln(x)]}{s^2}\right]^{\eps_{max}}_{\eps_{min}}
\end{aligned}
.\end{equation}

% The computation of the emission photon rate in the Klein-Nishina regime is a bit more complex and we need to introduce several functions:

% \begin{equation}
% g_{th}(\eps_1, \eps) = \frac{N_e}{4 \bar{\gamma}^3} \: \eps_1^{1/2} \: \eps^{-3/2} \: \exp \left( - \sqrt{\frac{\eps_1}{\eps \bar{\gamma}^2}} \right)
% \end{equation}

% \begin{equation}
% g_{kn} \left( \eps_1, \eps\right) = \frac{3N_e}{16\bar{\gamma}^3} \frac{\ln(2\eps_1\eps) +1/2 }{\eps} \eps_1 \exp(-\eps_1/\bar{\gamma})
% \end{equation}

%\begin{equation}
%       K_1^{(s)} (\eps_{min}, \eps_{max}) = \left [ \frac{x^{-s}}{s} \right]^{\eps_{max}}_{\eps_{min}} 
%\end{equation}
%
%\begin{equation}
%       K_{2}^{(s)}(\eps_{min},\eps_{max}) = \left [ \frac{x \exp(-(s+1)\ln(x))}{s^{2}}   (1+ s \ln(x)) \right ]^{\eps_{max}}_{\eps_{min}} 
% \end{equation}

As shown before, the synchrotron spectrum resulting from a pile-up distribution is a broken power law, $\displaystyle \d n_{ph}/\d \eps \d t = n_s \eps^{-s}$ , with $s=-1$ in the optically thick part $[\eps_{min}, \eps_{abs}]$ and $s=2/3$ in the optically thin part $[\eps_{abs}, \eps_{max}]$. One must therefore consider two cases depending on the position of the absorption energy $\eps_{abs}$ relative to the Thomson/Klein-Nishina threshold $1/\eps_1$. The SSC spectrum in the Klein-Nishina regime is finally given by
\begin{equation}
\label{eq:ssc_spectrum_final_appendix}
\frac{\d n^{kn}(\eps_1)}{\d \eps_1 \d t} = n_0 n_s c \sigma_{Th} \left( J_{Th}(\eps_1) + J_{kn}(\eps_1) \right)
.\end{equation}

\begin{itemize}
\item
If $\eps_{abs} < 1/\eps_1,$

\begin{equation}
\begin{aligned}
        & \frac{\d n^{kn}(\eps_{1}) }{\d \eps_{1} \d t}  = c \sigma_{Th} \left[
                        \int_{\eps_{min}}^{\eps_{abs}} \d \eps \frac{\d n_{{ph}}}{\d \eps \d t} g_{th}(\eps_{1},\eps) \right. \\ 
                        & \left. +  \int_{\eps_{abs}}^{1/\eps_{1}} \d \eps \frac{\d n_{{ph}}}{\d \eps \d t} g_{th}(\eps_{1},\eps)          
                        +  \int_{1/\eps_{1}}^{\eps_{max}} \d \eps \frac{\d n_{{ph}}}{\d \eps \d t} g_{kn}(\eps_{1},\eps)
        \right]
\end{aligned}
,\end{equation}

which gives
\begin{equation}
\label{eq:ssc_kn_th_absinf}
\begin{aligned}
J_{Th}(\eps_1) = \eps_1 \bar{\gamma}^{-1} H_1^{(-1)} (\eps_{min}, \eps_{abs})
+ \eps_1^{-2/3} \bar{\gamma}^{7/3} H_1^{(2/3)} (\eps_{abs}, 1/\eps_{1})
\end{aligned}
,\end{equation}
\begin{equation}
\label{eq:ssc_kn_kn_absinf}
\begin{aligned}
J_{KN}(\eps_1) & = \frac{3}{8} \eps_1 \exp\left(-\frac{\eps_1}{\bar{\gamma}}\right)
\left\{ \frac{}{} K_2^{(2/3)}(1/\eps_1, \eps_{max}) \right. \\
& + \left. \left(\ln(2\eps_1) +\frac{1}{2} \right) K_1^{(2/3)}(1/\eps_1, \eps_{max})   \right\} 
\end{aligned}
.\end{equation}

%\begin{equation}
%       \mathcal{ J }_{th}(\eps_{1}) = \eps_{1} \bar{\gamma} 
%\end{equation}

\item
If $\eps_{abs} > 1/\eps_1,$

\begin{equation}
\begin{aligned}
        & \frac{\d n^{kn}(\eps_{1}) }{\d \eps_{1} \d t}  = c \sigma_{Th} \left[
                        \int_{\eps_{min}}^{1/\eps_{1}} \d \eps \frac{\d n_{{ph}}}{\d \eps \d t} g_{th}(\eps_{1},\eps) \right. \\
                        & + \left. \int_{1/\eps_{1}}^{\eps_{abs}} \d \eps \frac{\d n_{{ph}}}{\d \eps \d t} g_{kn}(\eps_{1},\eps)          
                        +  \int_{\eps_{abs}}^{\eps_0} \d \eps \frac{\d n_{ph}}{\d \eps \d t} g_{kn}(\eps_{1},\eps)
        \right]
\end{aligned}
,\end{equation}

which gives
\begin{equation}
\label{eq:ssc_kn_th_abssup}
J_{Th}(\eps_1) = \eps_1 \bar{\gamma}^{-1} H_1^{(-1)} (\eps_{min}, 1/\eps_1)
,\end{equation}
\begin{equation}
\label{eq:ssc_kn_kn_abssup}
\begin{aligned}
& J_{KN}(\eps_1) = \frac{3}{8} \eps_1 \exp\left(-\frac{\eps_1}{\bar{\gamma}}\right) \\
& \times \left\{ \left(\ln(2\eps_1) +\frac{1}{2} \right) \left( K_1^{(-1)}(1/\eps_1, \eps_{abs}) +  K_2^{(-1)}(1/\eps_1, \eps_{abs})\right) \right.\\
& \left. + K_1^{(2/3)}(\eps_{abs}, \eps_{max}) +  K_2^{(2/3)}(\eps_{abs}, \eps_{max}) \frac{}{}\right\} 
\end{aligned}
.\end{equation}

\end{itemize}

Finally, the continuity between the two regimes is done with the following connection.
\begin{equation}
        \cfrac{\d n_{ssc}(\eps_1)}{\d \eps_1 \d t} = \cfrac{ \cfrac{\d n^{Th}(\eps_1)}{\d \eps_1 \d t} + x^n \cfrac{\d n^{kn}(\eps_1)}{\d \eps_1 \d t} }{1 + x^n}
        \qquad \text{with} \quad x = \eps_{1}\eps_0
.\end{equation}

\section{Computation of the inverse-Compton scattering over a thermal distribution of photons \label{sec:appendix_ec}}

The number of scattered photons of energy $\eps'_{1} = h \nu'_{1}/m_{e}c^{2}$ per element of time, per energy, per energy of the incoming photon $\eps'$ and per elementary solid angle $\di \left( \frac{\d N}{\d t' \d \nu' \d \nu'_{1} \d \Omega'_{1} } \right)$ is equal to the number of photons $\di \left( \frac{I'_{\nu'}}{h\nu'} \right)$ at the energy $\eps' = h\nu'$ verifying $\di \epsilon'_1 = \frac{\epsilon'}{1 + \epsilon' (1-\cos \Phi')}$ multiplied by their probability to scatter in an emission solid angle $\d \Omega'_{1}$ (see \citealt{1968PhRv..167.1159J}):

\begin{align}
\frac{\d N'_{ec}}{\d t' \d \epsilon'_1 \d \epsilon' \d \Omega'_1}  = \int   &   \left(\frac{\d \sigma}{\d \Omega_1}\right)'   \frac{1}{h} \frac{I'_{\nu}}{m_e c^2 \epsilon'} \nonumber \\ 
& \times \delta\left( \eps'_{1} - \frac{\epsilon'}{1 + \epsilon' (1-\cos \Phi')}\right)\d \Omega'_{i}. 
\label{eq:compton_spectrum}
\end{align}

The distribution of photons, given by $\di I_\nu$ , should be given by Planck's law in the case of a radiative black body.
The approximation that we propose for the inverse-Compton scattering is based on an approximation of Planck's law that we call modified Wien's law:

% \begin{equation}
% \label{eq:wien_law}
%       W_{\nu}(\nu,T) = A(T) \, \nu^{2} \exp\left(-\frac{\nu}{\bar{\nu}(T)} \right) %\, \delta \left(\Omega - \Omega_{0}\right)
% \end{equation}
\begin{equation}
\label{eq:wien_law}
        W_{\nu}(\eps,T) = A(T) \, \eps^{2} \exp\left(-\frac{\eps}{\bar{\eps}(T)} \right) %\, \delta \left(\Omega - \Omega_{0}\right)
    % \left( \frac{m_e c^2}{h}\right)^2
,\end{equation}

leaving the parameters $A(T)$ and $\bar{\eps}(T)$  to be determined.

One can see the modified Wien's law as a mix of Rayleigh-Jeans law and of Wien's law.
An (arbitrary) criteria to determine $A(T)$ and $\bar{\eps}(T)$ is to conserve the total emitted power and the position of the emission peak in the Thomson regime. In this case, one can show that
% \begin{align}
% \label{eq:Wien_nu_A}
% &     \bar{\nu}(T) = \frac{\alpha}{2h} k_{B} \, T
% &     A (T) = \frac{2 \sigma h^{2}}{\pi k_{B}^{2} \alpha^{2}} T^{2}
% \end{align}
\begin{equation}
\label{eq:Wien_nu_A}
\begin{aligned}
&       \bar{\eps}(T) = \frac{\alpha k_{B}}{2h} \frac{m_e c^2}{h} \, T
&       A (T) = \frac{2 \sigma}{\pi} \left( \frac{m_e c^2}{k_{B} \alpha} \right)^2 T^{2}
\end{aligned}
,\end{equation}
with $\displaystyle \alpha \approx 2.821$ being a numerical constant.

% \corr{supprimer cette approx ?!!}

% Here we chose to conserve: 
% \begin{itemize}
%       \item the emitted power after scattering in the Thomson regime
%       \item the peak position of the emitted spectra in the Thomson regime
% \end{itemize}

% This leads to the resulting values:
% \begin{align}
% & \bar{\eps} = 1.31 \frac{k_B T}{m_e c^2}  & A = 2.90 \frac{k_B T}{ c^2}
% \end{align}

When replacing $\di I_\nu$ with $\di W_\nu$ in equation \ref{eq:compton_spectrum} and using the highly relativistic approximation $(\gamma \gg 1)$ and the head-on approximation \citep{BLUMENTHAL:1970gb}, one can show that the number of inverse-Compton photons scattered in the particle rest frame per energy unit and per time unit (denoted by $'$ subscript) for a thermal distribution of soft photons of energy $\gamma m_e c^2$ is given by

\begin{equation}
\label{eq:ic_wien_final_appendix}
   \begin{aligned}
        \frac{\d N_{ec}(\eps_1)}{\d t \d \eps_1} = & \frac{m^2_e c^4 r^2_e A \epsilon' \pi}{h^3 \gamma^4 (1-\beta\cos\theta_0)} \frac{1}{(1-x)} \left\lbrace \left( \frac{2}{\mathcal{H}} + 2 + x\bar{\epsilon}'\right) e^{-\mathcal{H}/2} \right.\\
          & \left. - \frac{}{} \left(2+\mathcal{H}\right) \left(E_1(\mathcal{H}/2)-E_1(2\gamma^2 \mathcal{H})\right) \right\rbrace
   \end{aligned}
,\end{equation}

with $\di x = \frac{\eps_1}{\gamma}$, $\theta_0$ being the angle between the incoming photon and the particle direction of motion,
$\di \mathcal{H}=\frac{x}{\bar{\epsilon}'(1-x)}  $
and $\di E_1(x) = \int_x^\infty \frac{\exp(-t)}{t} \d t$ being the exponential integral that can be easily computed by any numerical package.

\subsection{Approximations in the Thomson regime}

In the Thomson regime, the expression (\ref{eq:ic_wien_final_appendix}) can be simplified, especially with $\di x  \ll 1$ and $\di x \eps'_1  \ll  \frac{2 x}{\eps'_1} \approx \frac{2}{\mathcal{H}} $ and one obtains

\begin{equation}
\label{eq:ec_thomson}
\begin{aligned}
         \left(\frac{\d N_{ec}(\eps_1)}{\d t \d \eps_1} \right)_{Th} = & \frac{m^2_e c^4 r^2_e A \epsilon' \pi}{h^3 \gamma^4 (1-\beta\cos\theta_0)}  \left\lbrace \left( \frac{2}{\mathcal{H}} + 2 \right) \exp(-\mathcal{H}/2) \right.  \\
         & - \left. \frac{}{} (2+\mathcal{H}) E_1(\mathcal{H}/2)\right\rbrace
\end{aligned}
.\end{equation}
Therefore, the spectrum of the inverse-Compton scattering of a thermal distribution of photons on a single particle in the Thomson regime is a function of a single variable. 
%To fasten the computation in the Thomson regime, a tabulated version of this single variable function can be created.
%We can compute this function once, tabulate its values and interpolate on them when required. This way, the computation of the inverse Compton spectra is much faster. 

\subsection*{\label{sec:thomson_pileup}Thomson regime and pile-up particle energy distribution}

The kernel computed above can then be integrated over a particle distribution.
In the case of a pile-up distribution $\di n_{e}(\gamma) = \frac{N_{e}}{2\bar{\gamma}^{3}}\gamma^{2}\exp\left(-\gamma/\bar{\gamma}\right)$ this integration simplifies and one gets directly the inverse-Compton spectrum:

%The integration over $\gamma$ writes:
%
%\begin{equation}
%\frac{\d N_{1}}{\d t \d \eps_{1}} = \frac{\pi m_{e}^{2}c^{4}r_{e}^{2}A\eps_{1} }{h^{3}} \int_{1}^{\infty} N_{e} \frac{1}{\gamma^{2}(1-\beta\mu_{0})} f_{Th}(H) \exp\left( -\gamma/\bar{\gamma}\right) \d \gamma
%\end{equation}
%
%With the substitution $\displaystyle u = \gamma/\bar{\gamma}$ and the approximation $\gamma \gg 1$, one gets:

\begin{equation}
\label{eq:ec_pileup_thomson_appendix}
\frac{\d N_{ec}}{\d t \d \eps_{1}} = \frac{\pi m_{e}^{2}c^{4}r_{e}^{2}A\eps_{1} }{h^{3}}  \frac{N_e} {2\bar{\gamma}^4 (1-\mu_0) } \:  \chi(s)
,\end{equation}

with 
\begin{equation}
\chi(s) =   \int_0^\infty \frac{e^{-u}}{u^{2}} 
 \left\lbrace \left( \frac{2u^2}{s} + 2 \right) \exp\left(-\frac{s}{2u^2}\right) 
         - \left( 2+ \frac{s}{u^2} \right) E_1 \left(\frac{s}{2u^2}\right) \right\rbrace
\d u 
.\end{equation}

As $\chi(s)$ is a single variable function, it can be computed once, and then tabulated and interpolated over when required. This way, the computation of the inverse-Compton spectra from the scattering of a thermal soft photon field on a pile-up distribution of electrons can be done much faster than usually with complete numerical integration.

\subsection{Validating the approximation}

Here we compare our analytical approximation with a complete numerical integration of the inverse-Compton spectrum. The result is displayed in figure \ref{fig:wien-vs-ZP13}.

\begin{figure}[h]
\begin{center}
        \includegraphics[width=\hsize]{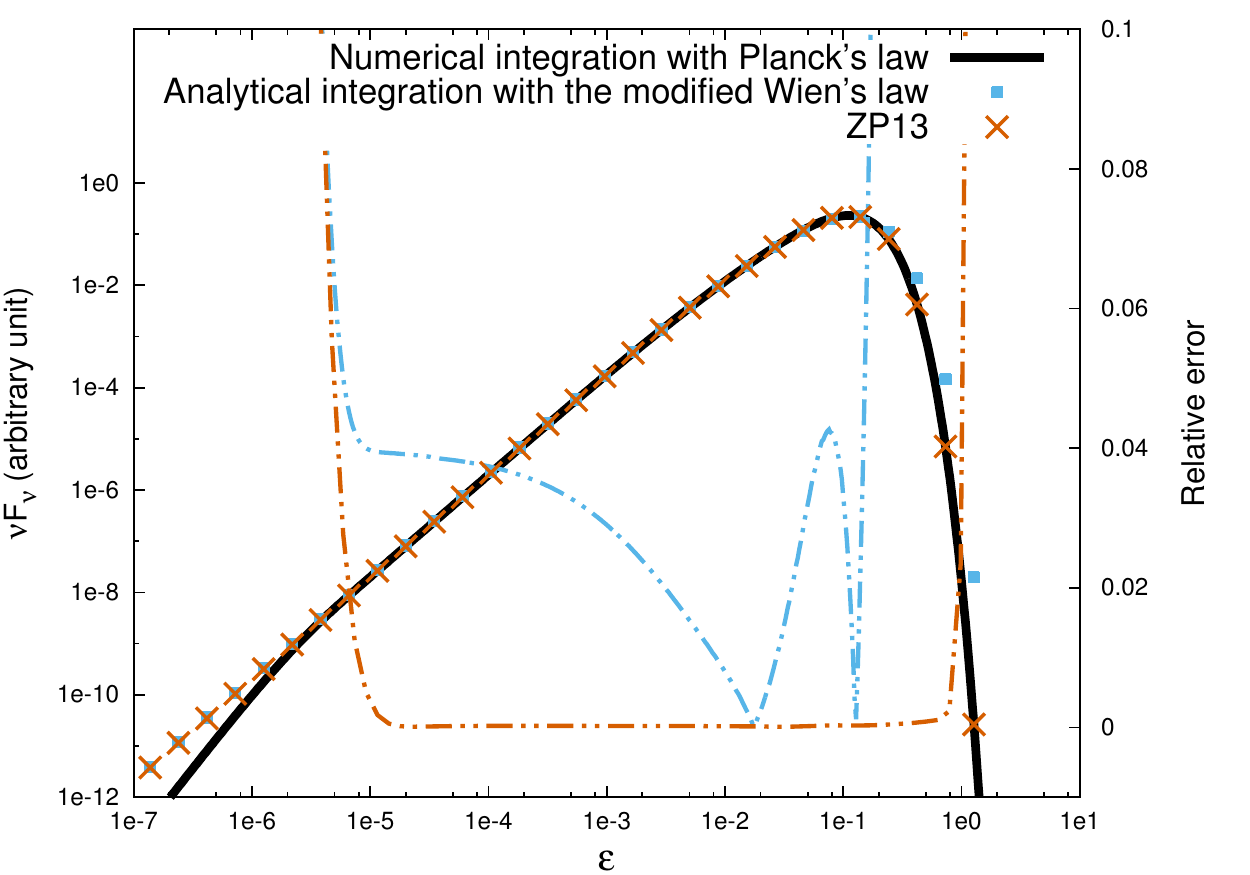}
\caption{Comparison of the different inverse-Compton spectra of a thermal soft photon field (T=5780K) scattering on a mono-energetic particle distribution of Lorentz factor $\gamma = 1e2$ with the relative errors in dot-dot-dashed thin lines. The numerically integrated spectra with the thermal distribution described by Planck's law is shown in black. The analytically integrated spectra with the thermal distribution described by the modified Wien's law (given by equation \ref{eq:ic_wien_final}) is shown in blue. As a comparison, we also provide the spectrum from \cite{Zdziarski:2013ed} in orange.}
\label{fig:wien-vs-ZP13}
\end{center}
\end{figure}

We see that the solution proposed by \cite{Zdziarski:2013ed} (hereafter ZP13) is, overall, better than ours with a relative error below 0.1\% in the medium range. The emission peak is also closer to the complete calculation.

Nevertheless, our calculation presents the advantage of being faster than the one from ZP13; in our simulations it was shown to be more than twice as fast in the case studied here. Moreover, another large reduction in computation time is also achieved with the one variable function $\chi(s)$ in the case of a pile-up distribution in the Thomson regime. We note that the error made compared to the exact numerical integration will be smoothed in the complete model as the spectra are convolved on several parameters and on the integration along the jet.
\end{appendix}

%The integral of $B_{\nu}(\nu,T)$ is well known and its result is given by Stefan-Boltzmann law:
%\begin{equation}
%\label{eq:Boltzmann}
%\iint  B_{\nu}(\nu,T) \d \nu \cos(\theta) \d \Omega = \sigma T^{4}
%\end{equation}

%And $\nu_{P}^{max}$ is given by Wien's displacement law:
%\begin{equation}
%\label{eq:nuPmax}
%\nu_{P}^{max} = \frac{\alpha}{h} k_{B} T \approx \left( 5.879 \times 10^{10} \, \text{Hz}\, K^{-1} \right) \, T
%\end{equation}
%with $\displaystyle \alpha \approx 2.821$ a numerical constant.

\end{document}